\documentclass[jmp,aps,showpacs,reprint]{revtex4-1}
\usepackage{graphicx}
\usepackage{dcolumn}
\usepackage{amsmath}
\usepackage{xcolor}
\usepackage{epsfig}
\RequirePackage{xspace}
\usepackage{threeparttable}

\usepackage{relsize}
\usepackage{hyperref}

\begin{document}

\title{
\large \bfseries \boldmath Study of $B\to  \pi  \rho$, $\pi \omega$ decays in the modified perturbative QCD approach}

\author{Sheng L\"{u}} \email{shenglyu@mail.nankai.edu.cn}
\author{Mao-Zhi Yang}\email{yangmz@nankai.edu.cn}
\affiliation{School of Physics, Nankai University, Tianjin 300071, People's Republic of China}

\date{\today}


\begin{abstract}
We investigate the $B\to\pi\rho$ and $\pi\omega$ decay processes using the modified perturbative QCD (PQCD) approach. Sudakov factors arising from the resummation of the double logarithms for the contributions of the transverse momenta of partons can increase the applicability of  perturbative calculations. In this work we find that infrared contributions cannot be fully suppressed by Sudakov factors alone when the $B$ meson wave function obtained by solving the wave equation in the QCD-inspired relativistic potential model is used in the calculation of $B$ decays. Therefore, it is necessary to introduce a soft cutoff scale  $\mu_c$ to separate the hard and soft contributions. Color-octet contributions for the intermediate quark-antiquark pairs are considered. Such contributions are crucial to deminish the discrepancies between the theoretical calculation and experiemntal data. Using the isospin symmetry, we do find the optimal parameter space for the nonperturbative inputs, which can result in excellent agreement between the theoretical calculation and experimental data of $B\to\pi\rho$ and $B\to\pi\omega$ decays.
\end{abstract}

\maketitle
\section{Introduction}

Weak decays of $B$ mesons exhibit rich phenomenology, and the study of them plays a crucial role in precision tests of the Standard Model (SM), understanding non-perturbative QCD effects, and searching for New Physics (NP). With the recent upgrade of Belle II and the start of its data recording, the expected accumulated statistics will reach 50 times of the combined total amount of the previous two $B$-factory experiments \cite{PDG2024,Belle2019}. Meanwhile, the LHCb experiment is also significantly improving its detection efficiency. In the coming years, we will be able to have precise measurements of numerous decay channels in rare $B$ decay processes. These developments will help to identify the limitations of current theoretical models, impose stringent constraints on the new components introduced in the theoretical frameworks and address urgent need to improve the precision of theoretical calculations.

After decades of scientific research, three main theoretical approaches have been developed, which are  proven to be mainly successful in describing $B$-meson decays. They are perturbative QCD (PQCD) approach \cite{PQCD1,PQCD2,PQCD3}, QCD factorization (QCDF) \cite{QCDf1,QCDf2,QCDf3,QCDf4}, and soft-collinear effective theory (SCET) \cite{SCET1,SCET2,SCET3,SCET4,SCET5,SCET6}. Beside the success of these theoretical methods in explaining experimental data in $B$ decays, they have to continuously face challenges coming from the more and more accurate experimental results, particularly for the branching fraction of $B^0 \to \pi^0\pi^0$ decay and the CP violation measurements in $B \to K\pi$ channels. Many efforts have been made to solve the puzzles in $B\to \pi\pi$ and $K\pi$ decays within \cite{LiMiSa2005,Li-Mishima2011,Li-Mishima2014,bai2014revi,bai2014revi,LLX2016,xiao2022,cheng-chua2005,cheng-chua2009a,cheng-chua2009b,CSYL2014,chua2018} and beyond \cite{Bar2004,Bae2005,Arn2006,Kim2008,Bea2018,Dat2019} the Standard Model. Furthermore, theoretical frameworks that attempt to resolve the $\pi\pi$ puzzle generally cannot succeed in simultaneously solving the $B\to \pi\pi$ and $K\pi$ puzzles while keeping the constraint from the experimental measurement of $B^0\to \rho^0\rho^0 $ decay \cite{Li-Mishima2006}.

In $B\to\pi\rho$ and $B\to\pi\omega$ decays, the decay channels of $B^0\to\pi^0\rho^0$ and $B^0\to\pi^0\omega$ are similar color-suppressed decays as $B^0\to\pi^0\pi^0$. Recent studies employing various theoretical approaches have made significant progress by making theoretical predictions more closer to experimental data for these decays \cite{QCDf4,LuYang2002,RGLu2012,WangLu2008,xiao2022}. However, as more experimental data were accumulated and experimental precision improves, theoretical calculation shows deviation from data especially for CP violation observables. Notably, calculations based on PQCD approach for branching fraction of $B^0\to\pi^0\rho^0$ decay yield values significantly lower than experimental measurements. Furthermore, the latest experimental measurements show that the direct CP violation  for $B^+\to\pi^+\rho^0$ and $B^+\to\pi^0\rho^+$ channels are close to zero with high precision, while all theoretical approaches predict values significantly deviating from zero. These discrepancies indicate the necessity of performing further theoretical study for $B\to\pi\rho$ and $B\to\pi\omega $ decays.

Recently, we have developed an modified theoretical framework based on the PQCD approach \cite{Lu-Yang2021,lu-yang2023,wang-yang2023,WangYang2025,GuiYang2025}. Utilizing the $B$-meson wave function derived from the QCD-inspired relativistic potential model \cite{Yang2012,LY2014,LY2015,SY2017,SY2019}, we have identified non-negligible infrared contributions still existing in the decay amplitude inspite of the suppression effect of Sudakov factors. These residual infrared contributions require careful treatment. To remove these infrared contribution, we introduce a cutoff scale $\mu_c$ to separate perturbative ($\mu > \mu_c$) and non-perturbative ($\mu < \mu_c$) regimes. By incorporating soft form factors and color-octet state contributions, our mothod can resolve both the $\pi\pi$ and $K\pi$ puzzles simultaneously and explain all the measured branching fractions and CP-violation parameters for $B \to PP$ decays, where $P$ denotes a pseudoscalar meson.

In this work, we extend our theoretical framework to investigate the $B\to\pi\rho$ and $B\to\pi\omega$ decay processes, where the final state contains one vector meson. Our approach retains transverse momentum in perturbative calculations and includes dominant next-to-leading order (NLO) contributions. We introduce a cutoff scale $\mu_c$ to separate perturbative and non-perturbative regimes, with soft transition form factors and soft production form factors accounting for infrared contributions below $\mu_c$. The color-octet mechanism is consistently incorporated in our framework. Through global fitting, we obtain an optimal parameter space that yields a reduced $\chi^2$ value of 1.18, demonstrating excellent agreement between our theoretical predictions and experimental measurements.

The organization of the remaining part of this paper is as follows: Sec.~II briefly introduces the theoretical framework of the PQCD approach. We present the calculations of leading-order (LO) and next-to-leading-order (NLO) perturbative contributions in section~III and ~IV respectively. Sec.~V incorporates both soft form factors and color-octet contributions. Sec.~VI is devoted to present our numerical results and provides detailed analysis and discussion. Finally, Sec.~VII gives a brief summary.

\section{The framework}

\subsection{The Factorization Formula For The Decay Amplitude}

In the decay process of a $B$ meson into two light mesons, such as $B \to \pi \rho (\omega)$ decay, the two outgoing light mesons obtain momenta that are very large compared to their own masses, and there are two steps in this decay. The first one is the light quark emission process, where the heavy quark $b$ in the $B$ meson decays into three light quarks via weak interaction. The second one is the hadronization process, where the final quarks from the $b$ quark decay hadronize into light mesons together with the spectator quark in $B$ meson. In this step, the light spectator quark needs to be boosted into a state with large momentum by absorbing hard gluon. There are also contributions of quark-antiquark annihilation process, where the heavy $b$ quark and light antiquark in $B$ meson annihilate into a light qurk-antiquark pair via electroweak interaction, and then a pair of quark-antiquark is excited from the vacuum by hard gluon. In all of these cases, the dominant contribution arises from QCD processs involving hard gluon in general, which makes the application of perturbative QCD (PQCD) methods feasible. The application of this method is based on the proof of the PQCD factorization theorem \cite{Li1995,liyu1996-1,liyu1996-2}. According to the factorization theorem, the amplitude of the $B$ meson decay process can be expressed as a convolution of three parts

\begin{eqnarray}
&&\mathcal{M}=\int d^3k \int d^3k_1 \int d^3 k_2 \Phi^B(k,\mu)\nonumber \\
&&\times C(\mu)  H(k,k_1,k_2,\mu) \Phi^{M_1} (k_1,\mu)\Phi^{M_2} (k_2,\mu),
\end{eqnarray}
where $H(k,\mu)$ represents the hard amplitude, which is process-dependent and can be calculated perturbatively. The hard amplitude describes the short-distance interactions mediated by hard gluons, and it can be typically computed using perturbative QCD techniques. The symbols $\Phi^B(k,\mu), \Phi^{M_1} (k_1,\mu), $ and $\Phi^{M_2} (k_2,\mu) $
 are the wave functions of the $B$ meson and the two final-state light mesons, respectively. These wave functions contain the non-perturbative effects of the meson's internal interaction. These wave functions can be determined using non-perturbative methods such as QCD sum rules, potential model or extracted from experimental data. They describe the distribution of quarks and gluons inside the meson and are essential for connecting the short-distance hard interactions with the long-distance hadronization process. Finally, $C(\mu)$ represents the Wilson coefficients, which encode the effects of the renormalization-group evolution from the electroweak scale to the scale of the $B$ meson decay. These coefficients are calculated using the operator product expansion (OPE) and are process-independent, depending only on the underlying weak interaction operators.

 In non-leptonic decays of $B$ mesons, three characteristic scales are involved: the mass of the $W$ boson $m_W$, the characteristic scale of the decay dynamics $t$, and the factorization scale $1/b$, where $b$ is the conjugate variable of the parton transverse momentum. Radiative corrections give rise to two types of large logarithmic terms: $\ln(m_W/t)$ and $\ln(tb)$. By applying the renormalization group equation method, the evolution effects from the scale $m_W$ to $t$ are encapsulated in the Wilson coefficients, while the evolution effects from $1/b$ to $t$ yield a factor $\mathrm{exp}\left[-2\int_{1/b}^t\frac{d\mu}{\mu}\gamma_q(\alpha_s(\mu))\right]$.

 Additionally, there are two types of double logarithmic terms: $\ln^2(k_T/t)$ and $\ln^2 x$. Using the resummation method, the former gives rise to the Sudakov factor \cite{liyu1996-1,liyu1996-2}, and the latter one results in the threshold resummation factor \cite{lihn2002,TK-HNL}. The Sudakov factor significantly suppresses long-distance contributions, while the threshold resummation factor eliminates endpoint divergence behavior, thereby enhancing the effectiveness of the PQCD approach. The explicit analytical forms of the Sudakov factor and the threshold resummation factor are provided in Appendix A.

 \subsection{The Effective Hamiltonian}

 After resumming the large logarithmic terms, we are allowed to calculate the hard decay amplitude $H(k,\mu)$ within the framework of the four-quark effective theory. For the study of the $B \to \pi\rho$ decay process, we need to use the effective weak Hamiltonian for the $b \to d$ transition \cite{Hamiltanion1996},
\begin{eqnarray}
	\mathcal{H}_{\mathrm{eff}} &=& \frac{G_F}{\sqrt{2}}\bigg[ V_{ub}V_{ud}^*\big(C_1O_1^u+C_2 O_2^u\big)\nonumber \\
	&-&V_{tb}V_{td}^*\bigg(\sum_{i=3}^{10} C_i O_i+C_{8\textsl{g} } O_{8\textsl{g} } \bigg) \bigg],
\end{eqnarray}
where $G_F = 1.16638 \times 10^{-5}~\mathrm{GeV}^{-2}$ is the Fermi constant, $V_{ub}V_{ud}^*$ and $V_{tb}V_{td}^*$ represent the products of the Cabibbo-Kobayashi-Maskawa (CKM) matrix elements, $C_i$'s are the corresponding Wilson coefficients coefficients, and the operators are
\begin{eqnarray}
	&&O_1^u = \bar{q}_{\alpha}\gamma^{\mu}L u_{\beta}\cdot \bar{u}_{\beta}\gamma_{\mu}L b_{\alpha},
	 \nonumber  \\
	&&O_2^u = \bar{q}_{\alpha}\gamma^{\mu}L u_{\alpha}\cdot \bar{u}_{\beta}\gamma_{\mu}L b_{\beta},
	\nonumber  \\	
	&&O_3 = \bar{q}_{\alpha}\gamma^{\mu}L b_{\alpha}\cdot \sum_{q'}\bar{q}'_{\beta}\gamma_{\mu}L q'_{\beta},
	\nonumber  \\
	&&O_4 = \bar{q}_{\alpha}\gamma^{\mu}L b_{\beta}\cdot \sum_{q'}\bar{q}'_{\beta}\gamma_{\mu}L q'_{\alpha},
	\nonumber  \\	
	&&O_5 = \bar{q}_{\alpha}\gamma^{\mu}L b_{\alpha}\cdot \sum_{q'}\bar{q}'_{\beta}\gamma_{\mu}R q'_{\beta},
	\nonumber  \\	
	&&O_6 = \bar{q}_{\alpha}\gamma^{\mu}L b_{\beta}\cdot \sum_{q'}\bar{q}'_{\beta}\gamma_{\mu}R q'_{\alpha},
  \\	
	&&O_7 = \frac{3}{2}\bar{q}_{\alpha}\gamma^{\mu}L b_{\alpha}\cdot \sum_{q'} e_{q'} \bar{q}'_{\beta}\gamma_{\mu}R q'_{\beta},
	\nonumber  \\	
	&&O_8 =  \frac{3}{2}\bar{q}_{\alpha}\gamma^{\mu}L b_{\beta}\cdot \sum_{q'} e_{q'} \bar{q}'_{\beta}\gamma_{\mu}R q'_{\alpha},
	\nonumber  \\	
	&&O_9 =  \frac{3}{2}\bar{q}_{\alpha}\gamma^{\mu}L b_{\alpha}\cdot \sum_{q'} e_{q'} \bar{q}'_{\beta}\gamma_{\mu}L q'_{\beta}, \nonumber  \\
  	&&O_{10} =  \frac{3}{2}\bar{q}_{\alpha}\gamma^{\mu}L b_{\beta}\cdot \sum_{q'} e_{q'} \bar{q}'_{\beta}\gamma_{\mu}L q'_{\alpha},  	\nonumber \\
&& O_{8\textsl{g}}=\frac{g_s}{8\pi^2}m_b\bar{q}_\alpha\sigma^{\mu\nu}RT^a_{\alpha\beta}b_\beta G_{\mu\nu}^a, \nonumber	\end{eqnarray}
where $\alpha$ and $\beta$ are the color indices, and $L=(1-\gamma_5)$ and $R=(1+\gamma_5)$, which are the left- and right-handed projection operators. The sum relevant to $q'$ runs over all quark flavors being active at $m_b$ scale, that is $q'\in \{u,d,s,c,b\}$.

 \subsection{The Meson Wave Functions}

The meson wave function contains the momentum distribution information of partons within the bound state and absorbs non-perturbative effects, serving as a crucial input for calculating weak decays of $B$ meson.

For the $B$ meson, the spinor wave function can be defined using a non-local matrix element
$
\langle 0 | \bar{q}(z)_{\beta} [z, 0] b(0)_{\alpha} | B(p) \rangle,
$

\begin{equation}
\langle 0| \bar{q}(z)_\beta [z,0]b(0)_\alpha |\bar{B}\rangle =\int d^3k \Phi^{B}_{\alpha\beta}(\vec{k})e^{-ik\cdot z},
\end{equation}
where $\alpha$ and $\beta$ are spinor indices, and $[z, 0]$ is the path-ordered exponential (Wilson line), ensuring gauge invariance for the spinor wave function. Here, $\Phi$ represents the $B$ meson spinor wave function.

By solving the bound-state equation within the QCD-inspired relativistic potential model, the wave function of $B$ meson can be theoretically derived \cite{Yang2012,LY2014,LY2015}, where both the masses of $B$ meson and its higher excited states can be obtained in agreement with experimental data \cite{LY2014,LY2015}. With the wave function obtained in the potential model, the spinor wave function of $B$ meson can be derived, which is given by \cite{SY2017,SY2019}
\begin{eqnarray} \label{B-wave}
\Phi_{\alpha\beta}(&\vec{k}&)=\frac{-if_Bm_B}{4}K(\vec{k})
\nonumber\\
&& \cdot\Bigg\{(E_Q+m_Q)\frac{1+\not{v}}{2}\Bigg[\Bigg(\frac{k^+}{\sqrt{2}}  +\frac{m_q}{2}\Bigg)\not{n}_+
\nonumber\\
&&+\Bigg(\frac{k^-}{\sqrt{2}}  +\frac{m_q}{2}\Bigg)\not{n}_- -k_{\perp}^{\mu}\gamma_{\mu}  \Bigg]\gamma^5\nonumber\\
&&-(E_q+m_q)\frac{1-\not{v}}{2} \Bigg[  \Bigg(\frac{k^+}{\sqrt{2}}-\frac{m_q}{2}\Bigg)\not{n}_+
\nonumber\\
&& +\Bigg(\frac{k^-}{\sqrt{2}}-\frac{m_q}{2}\Bigg)\not{n}_--k_{\perp}^{\mu}\gamma_{\mu}\Bigg]\gamma^5
\Bigg\}_{\alpha\beta},\label{eqm}
\end{eqnarray}
where $k$ is the parton momentum, $f_B$ is the decay constant of the $B$ meson, $m_B$, $m_Q$, and $m_q$ are the masses of the $B$ meson, the heavy quark, and the light quark in the $B$ meson, respectively. $E_Q$ and $E_q$ are the energies of the heavy quark and the light quark. The four-velocity $v^\mu$ satisfies the relation:
\[
p_B^\mu = m_B v^\mu,
\]
where $p_B^\mu$ is the four-momentum of the $B$ meson. For convenience, we will adopt light-cone coordinates to describe the four-momentum in the following discussion. $n_\pm^\mu$ are two light-like vectors  $n_\pm^\mu=(1,0,0,\mp 1)$, and
\begin{equation}
k^\pm=\frac{E_q\pm k^3}{\sqrt{2}},\;\;\; k_\perp^\mu=(0,k^1,k^2,0). \label{kpm}
\end{equation}
The function $K(\vec{k})$ is proportional to the $B$ meson wave function:
\begin{equation}
K(\vec{k})=\frac{2N_B\Psi_0(\vec{k})}{\sqrt{E_qE_Q(E_q+m_q)(E_Q+m_Q)}} \label{wave-k}
\end{equation}
where $N_B$ is the normalization factor, defined as
\[
 N_B = \frac{i}{f_B}\sqrt{\frac{3}{(2\pi)^3 m_B}},
\]
 and $\Psi_0(\vec{k})$ is the $B$ meson wave function obtained by solving the wave equation in the rest frame of $B$ meson in the relativistic potential model, normalized as $\int d^3k |\Psi_0(\vec{k})|=1$. Its analytical form is given by \cite{SY2017}:
\begin{equation}\label{psi0}
\Psi_0(\vec{k})=a_1 e^{a_2|\vec{k}|^2+a_3|\vec{k}|+a_4},
\end{equation}
the parameters $a_i$ ($i=1,\cdots, 4$) are
\begin{eqnarray}
&&a_1=4.55_{-0.30}^{+0.40}\,\mathrm{GeV}^{-3/2},\quad\;
a_2=-0.39_{-0.20}^{+0.15}\,\mathrm{GeV}^{-2},\nonumber\\
&& a_3=-1.55\pm 0.20\,\mathrm{GeV}^{-1},\quad   a_4=-1.10_{-0.05}^{+0.10}.
\end{eqnarray}
Where the parameters $a_i$ ($i=1,\cdots, 4$) are obtained from fitting the numerical solution of the $B$ meson wave function derived from the relativistic potential model.

In the processes we study, the final-state mesons from $B$ meson decays are $\pi$ meson, $\rho$ meson, and $\omega$ meson. Since these are pseudoscalar and vector mesons, their wave functions are defined differently.

For the $\pi$ meson, the wave function in light-cone coordinates is given by\cite{bra1990,bal1999,Ball-Braun2006}
\begin{eqnarray}
\langle \pi(p_\pi)|\bar{q}(y)_\rho q'(0)_\delta |0\rangle &=&\int dx d^2k_{q\perp}e^{i(x p_\pi\cdot y-y_{\perp}\cdot k_{q\perp})}\nonumber\\
&&\times\Phi^\pi_{\delta\rho}
\end{eqnarray}
where $x$ is the longitudinal momentum fraction and $k_\perp$ is the transverse momentum.

The spinor wave function can be expanded in terms of twist-expansion as follows
\begin{eqnarray} \label{pion-spinor1}
\Phi^\pi_{\delta\rho}&=&\frac{if_\pi}{4}\Bigg\{\not{p}_\pi\gamma_5\phi_\pi(x,k_{q\perp})
-\mu_\pi\gamma_5\Bigg(\phi^\pi_P(x,k_{q\perp})\nonumber\\
&& -\sigma_{\mu\nu}p_\pi^\mu y^\nu\frac{\phi^\pi_\sigma(x,k_{q\perp})}{6}\Bigg)\Bigg\}_{\delta\rho}
\end{eqnarray}
where $f_\pi$ is the decay constant of the $\pi$ meson, $\mu_\pi = \frac{m^2_{\pi}}{m_u+m_d}$  is the chiral mass, and $\phi_\pi$, $\phi^\pi_P$ and $\phi^\pi_\sigma$ are distribution functions of twist-2 and twist-3. In momentum space, the expression of the spinor wave function becomes \cite{bf2001,wy2002}
\begin{eqnarray} \label{pion-spinor2}
\Phi^\pi_{\delta\rho}&=&\frac{if_\pi}{4}\Bigg\{\not{p}_\pi\gamma_5\phi_\pi(x,k_{q\perp})
-\mu_\pi\gamma_5\Bigg(\phi^\pi_P(x,k_{q\perp})\nonumber\\
&& -i\sigma_{\mu\nu}\frac{p_\pi^\mu \bar{p}_\pi^\nu}{p_\pi\cdot \bar{p}_\pi} \frac{\phi'^\pi_\sigma(x,k_{q\perp})}{6}\nonumber\\
&&+i  \sigma_{\mu\nu}p_\pi^\mu\frac{\phi^\pi_\sigma(x,k_{q\perp})}{6}\frac{\partial}{\partial k_{q\perp\nu}}\Bigg)\Bigg\}_{\delta\rho}
\end{eqnarray}
where $\bar{p}_\pi =(E_\pi,-\vec{p}_\pi)$ is the four-momentum vector with the direction opposite to that of the pion's motion, and $\phi'^{\pi}_\sigma (x,k_{q\perp})$ is the partial derivative of $\phi^\pi_\sigma (x,k_{q\perp})$ with respect to the longitudingal momentum fraction.

In the decay of $B \to \pi \rho$, due to angular momentum conservation, the $\rho$ meson is only longitudinally polarized. Therefore, we only consider the longitudinally polarized wave function \cite{TK-HNL,Ball-Braun1998}

\begin{eqnarray}
\langle \rho(p_{\rho},\epsilon_{L})|\bar{q}(y)_\gamma q'(0)_\delta |0\rangle &=&\int dx d^2k_{q\perp}e^{i(x p_\rho\cdot y-y_{\perp}\cdot k_{q\perp})}\nonumber\\
&&\times\Phi^{\rho}_{\delta\gamma},
\end{eqnarray}
where $\epsilon_L$ is the longitudinal polarization vector of the $\rho$ meson. The spinor wave function of the $\rho$ meson can be given in the following form
\begin{eqnarray}
\varPhi_{\delta\gamma}^{\rho} =&&\frac{i}{4}\Big\{\not{\epsilon}_L\Big[f_{\rho}^{||}m_{\rho}\phi_{\rho}(x,k_{q\perp})+f_{\rho}^{\perp}\not{p}_{\rho}\phi_{\rho}^{t}(x,k_{q\perp})\Big] \nonumber\\
&&+f_{\rho}^{\perp}m_\rho\phi_\rho^s(x,k_{q\perp})\Big\}_{\delta\gamma}
\end{eqnarray}
For longitudinally polarized vector mesons, there are two decay constants, $ f_{\rho}^{||} $ and $ f_{\rho}^{\perp} $. The twist-2 wave function is proportional to $ f_{\rho}^{||} $, while the twist-3 wave functions $\phi_{\rho}^{t}(x,k_{q\perp})$ and $\phi_\rho^s(x,k_{q\perp})$ are proportional to $f_{\rho}^{\perp} $. The wave functions $\phi_{\rho}(x,k_{q\perp}$, $\phi_{\rho}^{t}(x,k_{q\perp})$ and $\phi_\rho^s(x,k_{q\perp})$ can be found in Appendix B.

The wave function of $\omega$ meson is similar to that of the $\rho$ meson. One only needs to replace the relevant parameters with those of the $\omega$ meson.

\section{The Leading order contribution of the hard Amplitude }

Next, we calculate the hard scattering amplitude kernel $H(k,\mu)$, whose leading-order contribution involves the exchange of a hard gluon in the strong interaction. There are eight types of topological diagrams for the leading-order contributions, which are shown in Fig. 1. The subdiagrams (a), (b), (c), and (d) involve a hard gluon connecting the four-quark operator and the spectator quark, while (e), (f), (g), and (h) involve annihilation processes, where a hard gluon is emitted from one of the quarks in the four-quark operator, and excits a quark-antiquark pair from the vacuum.

\begin{widetext}
	
\begin{figure}[hbt]
	\begin{center}
		\epsfig{file=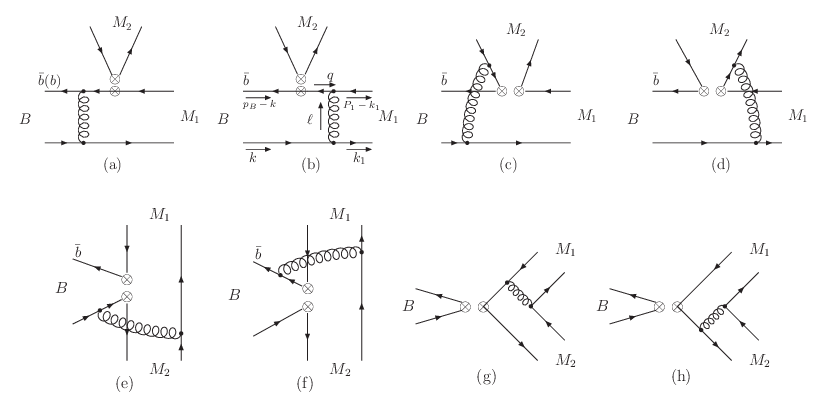,width=15cm,height=8cm}
		\caption{Diagrams contributing to the $B\rightarrow M_1 M_2$ decays at the leading order in QCD.} \label{fig1}
	\end{center}
\end{figure}

Diagrams (a) and (b) are factorizable because the part related to the light meson $M_2$ can be separated and directly traced, yielding the product of the meson decay constant and its momentum. For the case of $B \to \pi \rho$, where the externally emitted meson $M_2$ is the $\rho$ meson, the decay amplitude expression after including the Sudakov factor is given by
	\begin{eqnarray}\label{fe}
F_{e,\pi \rho} &=&-i\frac{4\pi^2}{N_c^2}f_B f_{\pi} f_{\rho}^{||} m_B\int dk_{\perp}k_{\perp}\int_{x^d}^{x^u}dx\int_0^1 dx_1 \int_0^\infty bdbb_1db_1 (\frac{1}{2}m_B+\frac{|\vec{k}_{\perp}|^2}{2x^2m_B})K(\vec{k})(E_Q+m_Q)\nonumber\\
&&\times J_0(k_{\perp}b) \Bigg\{\alpha_s(\mu_{e}^1)\Bigg(2m_B[E_q(1+x_1)+k^3(1-x_1)] \phi_{\pi}(\bar{x}_1,b_1)+2\mu_{\pi}[E_q(1-2x_1)-k^3]\phi^P_{\pi}(\bar{x}_1,b_1)\nonumber\\
&&-\frac{1}{3}\mu_{\pi}[E_q(1-2 x_1)-k^3]\phi'^\sigma_{\pi}(\bar{x}_1,b_1)\Bigg) h_e^1(x,x_1,b,b_1)S_t(x_1)\exp[-S_{B}(\mu_{e}^1)-S_{\pi}(\mu_{e}^1)] \nonumber\\
&&+\alpha_s(\mu_{e}^2) [4\mu_{\pi}(E_q-k^3)]\phi^P_{\pi}(\bar{x}_1,b_1)h_e^2(x,x_1,b,b_1) S_t(x)\exp[-S_{B}(\mu_{e}^2)-S_{\pi}(\mu_{e}^2)]\Bigg\}
\end{eqnarray}
where $N_c = 3$ is the number of color degrees of freedom, $x_i$'s are the momentum fractions of the antiquarks in the two light mesons, and $\bar{x}_i=1-x_i$, for $i=1,\; 2$. For the momentum fraction $x$ carried by the light quark in the $B$ meson, the integration limits are constrained by the requirement that the off-shell mass of the heavy quark is real in the relativistic potential model \cite{Lu-Yang2021}
\begin{equation}
x^{u,d}=\frac{1}{2}\pm \sqrt{\frac{1}{4}-\frac{|\vec{k_{\perp}}|^2}{m_B}}.
\end{equation}

During the calculation, we transform the integration over the transverse momentum $k_\perp$ from momentum space to $b$-space using a Fourier transform, where $b$ is the Fourier conjugate variable of $k_\perp$.

If the inserted operator is the current $(V-A)(V+A)$, for vector mesons, only the vector current contributes. Therefore,
\begin{equation}
F^R_{e,\pi \rho}=F_{e,\pi \rho}.
\end{equation}
The superscript $R$ indicates that the inserted current operator includes a right-handed current.

The $(V-A)(V+A)$ current, after Fierz transformation, becomes a $(S+P)(S-P)$ current, where $S$ denotes the scalar current and $P$ denotes the pseudoscalar current. Since both the scalar and pseudoscalar currents yield zero contribution when acting on the vector meson state, i.e. $\langle \rho | S+P | 0 \rangle = 0$, we get
\begin{equation}
F^P_{e,\pi \rho}=0.
\end{equation}
The superscript $P$ indicates that the inserted operator is the $(S+P)(S-P)$ current.

Diagrams (c) and (d) are non-factorizable. Their contributions are given by
\begin{eqnarray}\label{Me}
M_{e,\pi \rho} &=&-i\frac{4\pi^2}{N_c^2}f_B f_{\pi} f_{\rho}^{||} m_B\int dk_{\perp}k_{\perp}\int_{x^d}^{x^u}dx\int_0^1 dx_1 dx_2 \int_0^\infty bdbb_2db_2 (\frac{1}{2}m_B+\frac{|\vec{k}_{\perp}|^2}{2x^2m_B})K(\vec{k})(E_Q+m_Q)\nonumber\\
&&\times J_0(k_{\perp}b)\phi_{\rho}(\bar{x}_2,b_2) \Bigg\{\alpha_s(\mu_{d}^2)
\Bigg(-2m_B(x_2-1) (E_q+k^3)\phi_{\pi}(\bar{x}_1,b)-2\mu_{\pi} x_1(E_q-k^3)\phi^P_{\pi}(\bar{x}_1,b)\nonumber \\
&&-\frac{1}{3}\mu_{\pi} x_1(E_q-k^3)\phi'^\sigma_{\pi}(\bar{x}_1,b)\Bigg) h_d^1(x,x_1,x_2,b,b_2)\exp[-S_{B}(\mu_{d}^1)-S_{\pi}(\mu_{d}^1)-S_{\rho}(\mu_{d}^1)] \nonumber\\
&&+\alpha_s(\mu_{d}^2) \Bigg(-2m_B[E_q(x_1+x_2)+k^3(x_2-x_1) \phi_{\pi}(\bar{x}_1,b)+2\mu_{\pi} x_1(E_q+k^3)\phi^P_{\pi}(\bar{x}_1,b)\nonumber\\
&&-\frac{1}{3}\mu_{\pi} x_1(E_q+k^3)\phi'^\sigma_{\pi}(\bar{x}_1,b)\Bigg) h_d^2(x,x_1,x_2,b,b_2)\times \exp[-S_{B}(\mu_{d}^2)-S_{\pi}(\mu_{d}^2)-S_{\rho}(\mu_{d}^2)]\Bigg\}
\end{eqnarray}
for the insertion of operators of $(V-A)(V-A)$ current, and
\begin{eqnarray}\label{MeP}
M_{e,\pi \rho}^P &=&i\frac{4\pi^2}{N_c^2}f_B f_{\pi} f_{\rho}^{||} m_B\int dk_{\perp}k_{\perp}\int_{x^d}^{x^u}dx\int_0^1 dx_1 dx_2 \int_0^\infty bdbb_2db_2 (\frac{1}{2}m_B+\frac{|\vec{k}_{\perp}|^2}{2x^2m_B})K(\vec{k})(E_Q+m_Q)\nonumber\\
&&\times J_0(k_{\perp}b)\phi_{\rho}(\bar{x}_2,b_2) \Bigg\{\alpha_s(\mu_{d}^2)
\Bigg(-2m_B(E_q(x_1-x_2+1)-k^3(x_1+x_2-1))\phi_{\pi}(\bar{x}_1,b)\nonumber\\
&&+2\mu_{\pi} x_1(E_q+k^3)\phi^P_{\pi}(\bar{x}_1,b)-\frac{1}{3}\mu_{\pi} x_1(E_q+k^3) \phi'^\sigma_{\pi}(\bar{x}_1,b)\Bigg) h_d^1(x,x_1,x_2,b,b_2)\nonumber\\
&&\times \exp[-S_{B}(\mu_{d}^1)-S_{\pi}(\mu_{d}^1)-S_{\rho}(\mu_{d}^1)] +\alpha_s(\mu_{d}^2) \Bigg(2m_Bx_2(E_q+k^3)\phi_{\pi}(\bar{x}_1,b)\nonumber\\
&&-2\mu_{\pi} x_1(E_q-k^3)\phi^P_{\pi}(\bar{x}_1,b)-\frac{1}{3}\mu_{\pi} x_1(E_q-k^3)\phi'^\sigma_{\pi}(\bar{x}_1,b)\Bigg) h_d^2(x,x_1,x_2,b,b_2)\nonumber\\
&&\times\exp[-S_{B}(\mu_{d}^2)-S_{\pi}(\mu_{d}^2)-S_{\rho}(\mu_{d}^2)]\Bigg\}
\end{eqnarray}
for the Fierz transformed operators of $(S+P)(S-P)$ current. The following is
for contribution of the operators of $(V-A)(V+A)$ current
\begin{eqnarray}
M_{e,\pi \rho}^R&=&i\frac{4\pi^2}{N_c^2}f_Bf_{\pi} f_{\rho}^{\perp} \int dk_{\perp}k_{\perp}\int_{x^d}^{x^u}dx\int_0^1 dx_1 dx_2 \int_0^\infty bdbb_2db_2 (\frac{1}{2}m_B+\frac{|\vec{k}_{\perp}|^2}{2x^2m_B})K(\vec{k})(E_Q+m_Q)\nonumber\\
&&\times J_0(k_{\perp}b) \Bigg\{\alpha_s(\mu_{d}^2)
\Bigg[\frac{1}{3}m_{\rho} m_B(E_q+k^3)\phi_\pi(\bar{x}_1,b)\Big((x_2-1)\phi^t_{\rho}(\bar{x}_2,b_2) -6(x_2-1)\phi^s_{\rho}(\bar{x}_2,b_2)\Big) \nonumber
\end{eqnarray}
\begin{eqnarray}
&& +\frac{1}{3}\mu_{\pi} m_{\rho}\phi^P_{\pi}(\bar{x}_1,b)\Big([E_q(x_1+x_2-1)+k^3(-x_1 +x_2-1)]\phi^t_{\rho}(\bar{x}_2,b_2)+6[E_q(x_1-x_2+1)-k^3(x_1\nonumber\\
&&+x_2-1)]\phi^s_{\rho}(\bar{x}_2,b_2)\Big) -\frac{1}{18}\mu_{\pi}m_{\rho}\phi'^\pi_\sigma(\bar{x}_1,b) \Big( [E_q(x_1-x_2+1)-k^3(x_1+x_2-1)]\nonumber\\
&&\cdot \phi^t_{\rho}(\bar{x}_2,b_2) +6[E_q(x_1+x_2-1)+k^3(-x_1+x_2-1)]\phi^s_{\rho}(\bar{x}_2,b_2) \Big)\Bigg] h_d^1(x,x_1,x_2,b,b_2)\nonumber\\
&&\times\exp[-S_{B}(\mu_{d}^1)-S_{\pi}(\mu_{d}^1)-S_{\rho}(\mu_{d}^1)] +\alpha_s(\mu_{d}^2) \Bigg[-\frac{1}{3}m_{\rho}m_B(E_q+k^3)\Big(x_2\phi^t_{\rho}(\bar{x}_2,b_2)\nonumber\\
&& +6x_2\phi^s_{\rho}(\bar{x}_2,b_2)\Big)\phi_{\pi}(\bar{x}_1,b)+\frac{1}{3}\mu_{\pi} m_{\rho}\phi^P_{\pi}(\bar{x}_1,b)  \Big([E_q(x_1-x_2) -k^3(x_1+x_2)]\phi^t_{\rho}(\bar{x}_2,b_2)\nonumber\\
&& +6[E_q(x_1+x_2)+k^3(x_2-x_1)]\phi^s_{\rho}(\bar{x}_2,b_2)\Big) -\frac{1}{18}\mu_{\pi} m_{\rho}\phi'^\sigma_{\pi}(\bar{x}_1,b)\Big( [E_q(x_1+x_2)+k^3(x_2-x_1)]\nonumber\\
&&\cdot \phi^t_{\rho}(\bar{x}_2,b_2) +6[E_q(x_2-x_1)+k^3(x_1+x_2)] \cdot \phi^s_{\rho}(\bar{x}_2,b_2) \Big)\Bigg]h_d^2(x,x_1,x_2,b,b_2)\nonumber\\
&&\times\exp[-S_{B}(\mu_{d}^2)-S_{\pi}(\mu_{d}^2)-S_{\rho}(\mu_{d}^2)]\Bigg\}.
\end{eqnarray}

Fig. \ref{fig1} (e) and (f) are annihilation diagrams, and their contributions are generally small. The contribution for these diagrams is
\begin{eqnarray}\label{Ma}
M_{a,\pi \rho} &=&-i\frac{4\pi^2}{N_c^2 }f_B f_{\pi} \int dk_{\perp}k_{\perp}\int_{x^d}^{x^u}dx\int_0^1 dx_1 dx_2 \int_0^\infty bdbb_1db_1 (\frac{1}{2}m_B+\frac{|\vec{k}_{\perp}|^2}{2x^2m_B})K(\vec{k}) (E_Q+m_Q)\nonumber\\
&&\times J_0(k_{\perp}b)\Bigg\{\alpha_s(\mu_{f}^2)
\Bigg[-2m_B^2f_{\rho}^{||} (x_2-1)(E_q+k^3)\phi_{\pi}(\bar{x}_1,b_1)\phi_{\rho}(\bar{x}_2,b_1)+\frac{1}{3}\mu_{\pi}f_{\rho}^{\perp}m_{\rho} \phi^P_{\pi}(\bar{x}_1,b_1)\nonumber\\
&&\cdot\Big([-E_q(x_1+x_2-1)+k^3(x_1-x_2+1)]\phi^t_{\rho}(\bar{x}_2,b_1) +6[E_q(x_1-x_2+1)-k^3(x_1+x_2-1)]\phi^s_{\rho}(\bar{x}_2,b_1)\Big)\nonumber\\
&& -\frac{1}{18}\mu_{\pi}f_{\rho}^{\perp}m_{\rho} \phi'^\sigma_{\pi}(\bar{x}_1,b_1)\Big([E_q(x_1-x_2+1)-k^3(x_1 +x_2-1)]\phi^t_{\rho}(\bar{x}_2,b_1)-6[E_q(x_1+x_2-1)+k^3(-x_1\nonumber\\
&&+x_2-1)]\phi^s_{\rho}(\bar{x}_2,b_1) \Big)\Bigg]h_f^1(x,x_1,x_2,b,b_1)\exp[-S_{B}(\mu_{f}^1)-S_{\pi}(\mu_{f}^1)-S_{\rho}(\mu_{f}^1)]
\nonumber\\
&&+\alpha_s(\mu_{d}^2) \Bigg[-2m_B^2f_{\rho}^{||}x_1(E_q-k^3)\phi_{\pi}(\bar{x}_1,b_1)\phi_{\rho}(\bar{x}_2,b_1) -\frac{1}{3}\mu_{\pi}f_{\rho}^{\perp}m_{\rho}\phi^P_{\pi}(\bar{x}_1,b_1) \Big([E_q(x_1+x_2-1)+k^3(-x_1\nonumber\\
&&+x_2+1)]\phi^t_{\rho}(\bar{x}_2,b_1)+6[E_q(x_1-x_2+3)-k^3(x_1+x_2-1)]\phi^s_{\rho}(\bar{x}_2,b_1)\Big) +\frac{1}{18}\mu_{\pi}f_{\rho}^{\perp}m_{\rho}\phi'^\sigma_{\pi}(\bar{x}_1,b_1) \nonumber\\
&&\cdot\Big([E_q(x_1-x_2-1)-k^3(x_1+x_2-1)]\phi^t_{\rho}(\bar{x}_2,b_1) +6[E_q(x_1+x_2-1)+k^3(-x_1+x_2-3)]\phi^s_{\rho}(\bar{x}_2,b_1) \Big)\Bigg]\nonumber\\
&& \times h_f^2(x,x_1,x_2,b,b_1)\exp[-S_{B}(\mu_{f}^2)-S_{\pi}(\mu_{f}^2)-S_{\rho}(\mu_{f}^2)]\Bigg\}
\end{eqnarray}
for $(V-A)(V-A)$ current, and
\begin{eqnarray}
M_{a,\pi \rho} ^P &=&i\frac{4\pi^2}{N_c^2 }f_B f_{\pi}  \int dk_{\perp}k_{\perp}\int_{x^d}^{x^u}dx\int_0^1 dx_1 dx_2 \int_0^\infty bdbb_1db_1 (\frac{1}{2}m_B+\frac{|\vec{k}_{\perp}|^2}{2x^2m_B})K(\vec{k})(E_Q+m_Q)\nonumber\\
&&\times J_0(k_{\perp}b) \Bigg\{\alpha_s(\mu_{f}^2)\Bigg[2m_B^2f_{\rho}^{||}x_1(E_q-k^3)\phi_{\pi}(\bar{x}_1,b_1)\phi_{\rho}(\bar{x}_2,b_1)+\frac{1}{3}\mu_{\pi}f_{\rho}^{\perp}m_{\rho}\phi^P_{\pi}(\bar{x}_1,b_1)\Big([E_q(x_1+x_2-1)\nonumber\\
&&+k^3(-x_1+x_2-1)]\phi^t_{\rho}(\bar{x}_2,b_1)+6[E_q(x_1-x_2+1)-k^3(x_1+x_2-1)]\phi^s_{\rho}(\bar{x}_2,b_1)\Big) \nonumber\\
&& -\frac{1}{18}\mu_{\pi}f_{\rho}^{\perp}m_{\rho}\phi'^\sigma_{\pi}(\bar{x}_1,b_1)\Big([E_q(x_1-x_2+1) -k^3(x_1+x_2-1)]\phi^t_{\rho}(\bar{x}_2,b_1)  +6[E_q(x_1+x_2-1)\nonumber
\end{eqnarray}
\begin{eqnarray}\label{MaP}
&&+k^3(-x_1+x_2-1)]\phi^s_{\rho}(\bar{x}_2,b_1) \Big)\Bigg] h_f^1(x,x_1,x_2,b,b_1)\exp[-S_{B}(\mu_{f}^1)-S_{\pi}(\mu_{f}^1)-S_{\rho}(\mu_{f}^1)]\nonumber\\
&& +\alpha_s(\mu_{d}^2) \Bigg[2m_B^2f_{\rho}^{||}(x_2-1)(E_q+k^3)\phi_{\pi}(\bar{x}_1,b_1)\phi_{\rho}(\bar{x}_2,b_1) +\frac{1}{3}\mu_{\pi}f_{\rho}^{\perp}m_{\rho}\phi^P_{\pi}(\bar{x}_1,b_1) \nonumber\\
&&\Big( [E_q(x_1+x_2-1)+k^3(-x_1+x_2-3)]\phi'^\sigma_{M_2}(x_2,b_2-6[E_q(x_1-x_2+3)-k^3(x_1+x_2-1)]\phi^s_{\rho}(\bar{x}_2,b_1)\Big)\nonumber\\
&& +\frac{1}{18}\mu_{\pi}f_{\rho}^{\perp}m_{\rho}\phi'^\sigma_{\pi}(\bar{x}_1,b_1) \Big([E_q(x_1-x_2-1)-k^3(x_1+x_2-1)]\phi^t_{\rho}(\bar{x}_2,b_1) -6[E_q(x_1+x_2-1)+k^3(-x_1+x_2\nonumber\\
&&+1)]\phi^s_{\rho}(\bar{x}_2,b_1) \Big)\Bigg] h_f^2(x,x_1,x_2,b,b_1)\exp[-S_{B}(\mu_{f}^2)-S_{\pi}(\mu_{f}^2)-S_{\rho}(\mu_{f}^2)]\Bigg\}
\end{eqnarray}
for $(S+P)(S-P)$ current,
\begin{eqnarray}\label{MaR}
M_{a,\pi \rho} ^R &=&i\frac{4\pi^2}{N_c^2}f_B f_{\pi}  m_B\int dk_{\perp}k_{\perp}\int_{x^d}^{x^u}dx\int_0^1 dx_1 dx_2 \int_0^\infty bdbb_2db_2 (\frac{1}{2}m_B+\frac{|\vec{k}_{\perp}|^2}{2x^2m_B})K(\vec{k})(E_Q+m_Q)\nonumber\\
&&\times J_0(k_{\perp}b) \Bigg\{\alpha_s(\mu_{f}^2)\Bigg(\frac{1}{3}f_{\rho}^{\perp}m_{\rho}(E_q-k^3)\phi_{\pi}(\bar{x}_1,b_1)(x_2-1)\Big(\phi^t_{\rho}(\bar{x}_2,b_1) +6\phi^s_{\rho}(\bar{x}_2,b_1) \Big)\nonumber\\
&&-2f_{\rho}^{||} x_1\mu_{\pi}(E_q+k^3)\phi_{\rho}(\bar{x}_2,b_1)\phi^P_{\pi}(\bar{x}_1,b_1)+\frac{1}{3}f_{\rho}^{||} x_1\mu_{\pi}(E_q+k^3)\phi_{\rho}(\bar{x}_2,b_1)\phi'^\sigma_{\pi}(\bar{x}_1,b_1)\Bigg)\nonumber\\
&&\times h_f^1(x,x_1,x_2,b,b_2)\exp[-S_{B}(\mu_{f}^1)-S_{\pi}(\mu_{f}^1)-S_{\rho}(\mu_{f}^1)] +\alpha_s(\mu_{f}^2) \Bigg(-\frac{1}{3}f_{\rho}^{\perp}m_{\rho}[E_q(x_2+1)\nonumber\\
&&+k^3(x_2-1)]\Big(\phi^t_{\rho}(\bar{x}_2,b_1) +6\phi^s_{\rho}(\bar{x}_2,b_1)\Big)\phi_{\pi}(\bar{x}_1,b_1)-2f_{\rho}^{||} \mu_{\pi}[E_q(x_2-2)-k^3 x_2] \nonumber\\
&&\phi_{\rho}(\bar{x}_2,b_1)\phi^P_{\pi}(\bar{x}_1,b_1)+\frac{1}{3}f_{\rho}^{||}\mu_{\pi} [E_q(x_2-2)-k^3x_2]\phi_{\rho}(\bar{x}_2,b_1)\phi'^\sigma_{\pi}(\bar{x}_1,b_1)\Bigg) h_f^2(x,x_1,x_2,b,b_2)\nonumber\\
&& \times\exp[-S_{B}(\mu_{f}^2)-S_{\pi}(\mu_{f}^2)-S_{\rho}(\mu_{f}^2)]\Bigg\}
\end{eqnarray}
for $(V-A)(V+A)$ current.

Diagrams (g) and (h) are factorizable annihilation diagrams. For the insertion of the $(V-A)(V+A)$ operator, the contributions from diagrams (g) and (h) are approximately equal in magnitude but opposite in sign, leading to significant cancellation. However, due to differences in the wave functions of the final-state particles, a small residual contribution remains
\begin{eqnarray}\label{fa}
F_{a,\pi \rho}  &=&-i\frac{8\pi  }{N_c^2}f_B f_{\pi}  \int_0^1 dx_1 dx_2 \int_0^\infty b_1db_1b_2db_2
\Bigg\{\alpha_s(\mu_{a}^1)\Bigg(-m_B^2 (x_2-1) \phi_{\pi}(\bar{x}_1,b_1) f_{\rho}^{||}\phi_{\rho}(\bar{x}_2,b_2)\nonumber\\
&&-\frac{1}{3} \mu_{\pi} f_{\rho}^{\perp}m_{\rho}\Bigg[ x_2\phi^t_{\rho}(\bar{x}_2,b_2)+6(x_2-2)\phi^s_{\rho}(\bar{x}_2,b_2)\Bigg] \phi^P_{\pi}(\bar{x}_1,b_1)\Bigg)h_a^1(x_1,x_2,b_1,b_2)S_t(x_2) \exp[-S_{\pi}(\mu_{a}^1)\nonumber\\
&&-S_{\rho}(\mu_{a}^1)] +\alpha_s(\mu_{a}^2) \Bigg(-m_B^2 x_1\phi_{\pi}(\bar{x}_1,b_1) f_{\rho}^{||}\phi_{\rho}(\bar{x}_2,b_2)   -2\mu_{\pi} f_{\rho}^{\perp}m_{\rho}(x_2+1) \phi^P_{\pi}(\bar{x}_1,b_1)\phi^s_{\rho}(\bar{x}_2,b_2)\nonumber\\
&& +\frac{1}{3}\mu_{\pi} f_{\rho}^{\perp}m_{\rho}  (x_1-1)\phi'^\sigma_{\pi}(\bar{x}_1,b_1)\phi^s_{\rho}(\bar{x}_2,b_2)\Bigg) h_a^2(x_1,x_2,b_1,b_2)S_t(x_1)\exp[-S_{\pi}(\mu_{a}^2)-S_{\rho}(\mu_{a}^2)]\Bigg\}.
\end{eqnarray}

And
\begin{equation}
F_{a,\pi \rho}^R=-F_{a,\pi \rho} .
\end{equation}

The main contributions come from the operators of $(S+P)(S-P)$ currents. The result is
\begin{eqnarray}
F_{a,\pi \rho} ^P &=&i\frac{8\pi  }{N_c^2} \chi _B f_{\pi} \int_0^1 dx_1 dx_2 \int_0^\infty b_1db_1b_2db_2 \Bigg\{\alpha_s(\mu_{a}^1)\Bigg(-4\mu_{\pi}\phi^P_{\pi}(\bar{x}_1,b_1)f_{\rho}^{||}\phi_{\rho}(\bar{x}_2,b_2)\nonumber
\end{eqnarray}
\begin{eqnarray}\label{faP}
&&-\frac{1}{3}f_{\rho}^{\perp} m_{\rho}\Bigg[ (x_2-1)\phi^t_{\rho}(\bar{x}_2,b_2)-6(x_2-1)\phi^s_{\rho}(\bar{x}_2,b_2)\Bigg] \phi_{\pi}(\bar{x}_1,b_1)\Bigg)h_a^1(x_1,x_2,b_1,b_2)S_t(x_2) \exp[-S_{\pi}(\mu_{a}^1)\nonumber\\
&&-S_{\rho}(\mu_{a}^1)] +\alpha_s(\mu_{a}^2) \Bigg(-4m_{\rho}f_{\rho}^{\perp} \phi_{\pi}(\bar{x}_1,b_1)\phi^s_{\rho}(\bar{x}_2,b_2)-\frac{1}{3}\mu_{\pi} \Bigg[ x_1\phi'^\sigma_{\pi}(\bar{x}_1,b_1)+6x_1\phi^P_{\pi}(\bar{x}_1,b_1)\Bigg]f_{\rho}^{||}\phi_{\rho}(\bar{x}_2,b_2)\Bigg)\nonumber \\
&& \times h_a^2(x_1,x_2,b_1,b_2)S_t(x_1)\exp[-S_{\pi}(\mu_{a}^2)-S_{\rho}(\mu_{a}^2)]\Bigg\}.
\end{eqnarray}
Moreover, the contribution of annihilation diagram with $(S+P)(S-P)$ current contribute significantly to direct CP violation. In the above expression, $\chi_B$ arises from the matrix element of the $B$ meson with the insertion of the $(S+P)$ operator, i.e. $\langle 0 | S-P | B \rangle = -i\chi _B$, and
\begin{eqnarray} \label{chiB}
\chi _B &=& \pi f_B m_B \int dk_{\perp}k_{\perp}\int_{x^d}^{x^u}dx(\frac{1}{2}m_B+\frac{|\vec{k}_{\perp}|^2}{2x^2m_B})    K(\vec{k}) \Bigg[(E_q+m_q)(E_Q+m_Q)+(E_q^2-m_q^2)\Bigg].
\end{eqnarray}

For the decay process $B \to V_1 P_2$, it is necessary to interchange the corresponding meson masses and replace the meson wave functions as follows:

\begin{equation}
\begin{aligned}
&\phi_\pi(x_1) \rightarrow \phi_{\rho}(x_1),&
&\phi^P_\pi(x_1) \rightarrow -\phi^s_{\rho}(x_1),&
&\phi^{\sigma'}_\pi(x_1) \rightarrow- \phi^t_{\rho}(x_1),&
\end{aligned}
\end{equation}

\begin{equation}
\begin{aligned}
&\phi_\rho(x_2) \rightarrow \phi_{\pi}(x_2),&
&\phi^s_{\rho}(x_2) \rightarrow \phi^P_\pi(x_2),&
& \phi^t_{\rho}(x_2)\rightarrow \phi^{\sigma'}_\pi(x_2).&
\end{aligned}
\end{equation}

For $F^R_{a,V_1P_2}$, $F^P_{a,V_1P_2}$, and $M^R_{a,V_1P_2}$, the phase will also differ by a $\pi$ phase angle
	\begin{eqnarray}
\begin{aligned}
&F^R_{a,V_1P_2}\leftarrow - F^R_{a,P_1V_2},   &\qquad
F^P_{a,V_1P_2}\leftarrow - F^P_{a,P_1V_2},   & \qquad
&M^R_{a,V_1P_2}\leftarrow -M^R_{a,P_1V_2}. &
\end{aligned}
\end{eqnarray}

For diagrams (a) and (b) in Fig.~1, if the externally emitted meson is pion, the contribution from $F_{e,\rho \pi}^P $ is no longer zero. The result is given by
\begin{eqnarray}
F_{e,\rho \pi}^P &=&i\frac{4\pi^2}{N_c^2}f_B f_{\pi}  m_{\rho}\int dk_{\perp}k_{\perp}\int_{x^d}^{x^u}dx\int_0^1 dx_1 \int_0^\infty bdbb_1db_1 (\frac{1}{2}m_B+\frac{|\vec{k}_{\perp}|^2}{2x^2m_B})K(\vec{k})(E_Q+m_Q)\nonumber\\
&&\times J_0(k_{\perp}b) \Bigg\{\alpha_s(\mu_{e}^1)\Bigg(4m_B(E_q+k^3) f_{\rho}^{||}\phi_{\rho}(\bar{x}_1,b_1)-4m_{\rho}[E_q(x_1+2)-k^3 x_1] f_{\rho}^{\perp}\phi^s_{\rho}(\bar{x}_1,b_1)\nonumber\\
&&+\frac{2}{3}m_{\rho}[k^3(x_1-2)-E_q x_1] f_{\rho}^{\perp}\phi^t_{\rho}(\bar{x}_1,b_1)\Bigg) h_e^1(x,x_1,b,b_1) S_t(x_1)\exp[-S_{B}(\mu_{e}^1)-S_{\pi}(\mu_{e}^1)] \nonumber\\
&&-\alpha_s(\mu_{e}^2) [ 8m_{\rho}(E_q-k^3) ] f_{\rho}^{\perp}\phi^s_{\rho}(\bar{x}_1,b_1)h_e^2(x,x_1,b,b_1) S_t(x)\exp[-S_{B}(\mu_{e}^2)-S_{\rho}(\mu_{e}^2)]\Bigg\}.
\end{eqnarray}

In the above equations, the Sudakov factors $\exp[-S_{B}(\mu)]$, $\exp[-S_{\pi}(\mu)]$ and $\exp[-S_{\rho}(\mu)]$  are given in Appendix \ref{a}. $\phi^{(P,\sigma')}_{\pi}(x,b)$,  and $\phi^{(s,t)}_{\rho}(x,b)$ are the wave functions of light meson in $b$-space, which can be found in Appendix \ref{b}.

For the decay process $B \to \pi \omega$, one only needs to replace the relevant parameters of the $\rho$ meson with those of the $\omega$ meson. The functions $h_i$'s are given by

	\begin{eqnarray}
h_e^1(x,x_1,b,b_1)&=&K_0(\sqrt{xx_1}m_B b)\Big[\theta (b-b_1)I_0(\sqrt{x_1}m_B b_1)K_0(\sqrt{x_1}m_B b)
+\theta(b_1-b)\times I_0(\sqrt{x_1}m_B b)K_0(\sqrt{x_1}m_B b_1)\Big],\nonumber\\
&&
\end{eqnarray}
\begin{eqnarray}
h_e^2(x,x_1,b,b_1)&=&K_0(\sqrt{xx_1}m_B b)  \Big[\theta(b-b_1)I_0(\sqrt{x}m_B b_1)K_0(\sqrt{x}m_B b)+\theta(b_1-b) \times I_0(\sqrt{x}m_B b)K_0(\sqrt{x}m_B b_1)\Big],\nonumber\\
&&
\end{eqnarray}
\begin{eqnarray}
h_d^1(x,x_1,x_2,b,b_2)&=&K_0(-i\sqrt{x_1(1-x_2)}m_Bb_2)\Big[\theta(b_2-b)I_0(\sqrt{xx_1}m_B b)K_0(\sqrt{xx_1}m_Bb_2)+\theta(b-b_2) I_0(\sqrt{xx_1}m_Bb_2)\nonumber\\
&& \times K_0(\sqrt{xx_1}m_B b)\Big],
\end{eqnarray}
\begin{eqnarray}
h_d^2(x,x_1,x_2,b,b_2)&=&K_0(-i\sqrt{x_1x_2}m_B b_2)\Big[\theta(b_2-b)I_0(\sqrt{xx_1}m_B b)K_0(\sqrt{xx_1}m_B b_2)+\theta(b-b_2)I_0(\sqrt{xx_1}m_B b_2)\nonumber\\
&&\times K_0(\sqrt{xx_1}m_B b)\Big],
\end{eqnarray}
\begin{eqnarray}
h_f^1(x_1,x_2,b,b_1)&=&K_0(-i\sqrt{x_1(1-x_2)}m_B b)\Big[\theta(b-b_1)I_0(-i\sqrt{x_1(1-x_2)}m_B b_1)K_0(-i\sqrt{x_1(1-x_2)}m_B b)\nonumber\\
&&+\theta(b_1-b)I_0(-i\sqrt{x_1(1-x_2)}m_B b)K_0(-i\sqrt{x_1(1-x_2)}m_B b_1)\Big],
\end{eqnarray}
\begin{eqnarray}
h_f^2(x_1,x_2,b,b_1)&=&K_0(\sqrt{1-x_2+x_1x_2}m_Bb)\Big[\theta(b-b_1)I_0(-i\sqrt{x_1(1-x_2)}m_B b_1)K_0(-i\sqrt{x_1(1-x_2)}m_B b)\nonumber\\
&&+\theta(b_1-b)I_0(-i\sqrt{x_1(1-x_2)}m_B b)K_0(-i\sqrt{x_1(1-x_2)}m_B b_1)\Big],
\end{eqnarray}
\begin{eqnarray}
h_a^1(x_1,x_2,b_1,b_2)&=&K_0(-i\sqrt{x_1(1-x_2)}m_B b_1)\Big[\theta(b_2-b_1)I_0(-i\sqrt{1-x_2}m_B b_1)K_0(-i\sqrt{1-x_2}m_B b_2)\nonumber\\
&&+\theta(b_1-b_2)I_0(-i\sqrt{1-x_2}m_Bb_2)K_0(-i\sqrt{1-x_2}m_B b_1)\Big],
\end{eqnarray}
\begin{eqnarray}
h_a^2(x_1,x_2,b_1,b_2)&=&K_0(-i\sqrt{x_1(1-x_2)} b_1)\Big[\theta(b_2-b_1)I_0(-i\sqrt{x_1}m_B b_1)K_0(-i\sqrt{x_1}m_B b_2)+\theta(b_1-b_2)\nonumber\\
&&\cdot I_0(-i\sqrt{x_1}m_B b_2)K_0(-i\sqrt{x_1}m_B b_1)\Big].
\end{eqnarray}

For the renormalization scale $\mu$, we choose the maximum scale that appears in the calculation of the hard decay amplitude. This choice minimizes the logarithmic contributions in higher-order corrections, thereby reducing the impact of higher-order effects.

\begin{eqnarray}
\mu_e^1 &=& \max(\sqrt{x_1}m_B,\sqrt{xx_1}m_B,1/b,1/b_1),\nonumber\\
\mu_e^2 &=& \max(\sqrt{x}m_B,\sqrt{xx_1}m_B,1/b,1/b_1),\nonumber\\
\mu_d^1 &=& \max(\sqrt{xx_1}m_B,\sqrt{x_1(1-x_2)}m_B,1/b_1,1/b_2),\nonumber\\
\mu_d^2 &=& \max(\sqrt{xx_1}m_B,\sqrt{x_1x_2}m_B ,1/b_1,1/b_2),\nonumber\\
\mu_f^1 &=& \max(\sqrt{x_1(1-x_2)}m_B,1/b_1,1/b_2),\nonumber \\
\mu_f^2 &=& \max(\sqrt{x_1(1-x_2)}m_B,\sqrt{1-x_2+x_1x_2}m_B,  1/b_1,1/b_2),\nonumber\\
\mu_a^1 &=& \max(\sqrt{1-x_2}m_B,\sqrt{x_1(1-x_2)}m_B,1/b_1,1/b_2),\nonumber\\
\mu_a^2 &=& \max(\sqrt{x_1}m_B,\sqrt{x_1(1-x_2)}m_B,1/b_1,1/b_2).
\end{eqnarray}

Based on the quark composition of the initial and final-state mesons, we insert all the operators in the effective Hamiltonian in appropriate way to derive the final amplitudes for each decay channel

	\begin{eqnarray}\label{Mp+r-}
\mathcal{M}\bar{(B^0}\rightarrow\pi^+ \rho^-)=&&F_{e,\pi \rho}\Big[ \xi_{ud}(\frac{1}{3}C_1+C_2)-\xi_{td}(\frac{1}{3}C_3+C_4+\frac{1}{3}C_9+C_{10}) \Big]+M_{e,\pi \rho}\Big[ \xi_{ud}(\frac{1}{3}C_1)-\xi_{td}(\frac{1}{3}C_3+\frac{1}{3}C_9) \Big] \nonumber\\
&&+M^R_{e,\pi \rho}\Big[ -\xi_{td}(\frac{1}{3}C_5+\frac{1}{3}C_7) \Big]+M_{a, \pi \rho}\Big[ \xi_{ud}(\frac{1}{3}C_2)-\xi_{td}(\frac{1}{3}C_4+\frac{1}{3}C_{10}) \Big]\nonumber\\
&& +M_{a, \rho \pi}\Big[ -\xi_{td}(\frac{1}{3}C_3  +\frac{1}{3}C_4-\frac{1}{6}C_9-\frac{1}{6}C_{10}) \Big]+M^R_{a, \pi \rho}\Big[ -\xi_{td}(\frac{1}{3}C_5-\frac{1}{6}C_7) \Big]+M^P_{a, \pi\rho}\Big[ -\xi_{td}(\frac{1}{3}C_6\nonumber\\
&&+\frac{1}{3}C_{8}) \Big]+M^P_{a, \rho \pi}\Big[ -\xi_{td}(\frac{1}{3}C_6-\frac{1}{6}C_{8}) \Big]+F_{a,\pi\rho}\Big[\xi_{ud}(C_1+\frac{1}{3}C_2)-\xi_{td}(C_3+\frac{1}{3}C_4+C_9+\frac{1}{3}C_{10})\nonumber\\
&&+\xi_{td}(C_5+\frac{1}{3}C_6+C_7+\frac{1}{3}C_{8}) \Big]+F_{a,\rho\pi}\Big[-\xi_{td}(\frac{4}{3}C_3+\frac{4}{3}C_4-\frac{2}{3}C_9-\frac{2}{3}C_{10})+\xi_{td}(C_5+\frac{1}{3}C_6\nonumber\\
&&-\frac{1}{2}C_7-\frac{1}{6}C_{8}) \Big]+F^P_{a,\rho \pi }\Big[
-\xi_{td}(\frac{1}{3}C_5+C_6-\frac{1}{6}C_7-\frac{1}{2}C_8) \Big],
\end{eqnarray}	

\begin{eqnarray}\label{Mp+p-}
\mathcal{M}(\bar{B^0}\rightarrow\pi^- \rho^+)=&&F_{e,\rho \pi}\Big[ \xi_{ud}(\frac{1}{3}C_1+C_2)-\xi_{td}(\frac{1}{3}C_3+C_4+\frac{1}{3}C_9+C_{10}) \Big]-F^P_{e,\pi \rho} \Big[\xi_{td} (\frac{1}{3}C_5+C_6+\frac{1}{3}C_7+C_{8}) \Big]\nonumber\\
&&+M_{e,\rho \pi}\Big[ \xi_{ud}(\frac{1}{3}C_1)-\xi_{td}(\frac{1}{3}C_3+\frac{1}{3}C_9) \Big] +M^R_{e, \rho \pi}\Big[ -\xi_{td}(\frac{1}{3}C_5+\frac{1}{3}C_7) \Big]+M_{a,\rho \pi}\Big[ \xi_{ud}(\frac{1}{3}C_2)\nonumber\\
&&-\xi_{td}(\frac{1}{3}C_4+\frac{1}{3}C_{10}) \Big] +M_{a,\pi \rho}\Big[ -\xi_{td}(\frac{1}{3}C_3  +\frac{1}{3}C_4-\frac{1}{6}C_9-\frac{1}{6}C_{10}) \Big]+M^R_{a,\rho \pi}\Big[ -\xi_{td}(\frac{1}{3}C_5-\frac{1}{6}C_7) \Big]\nonumber\\
&&+M^P_{a,\rho \pi}\Big[ -\xi_{td}(\frac{1}{3}C_6+\frac{1}{3}C_{8}) \Big] +M^P_{a, \pi\rho}\Big[ -\xi_{td}(\frac{1}{3}C_6-\frac{1}{6}C_{8}) \Big]+F_{a,\rho \pi}\Big[\xi_{ud}(C_1+\frac{1}{3}C_2)\nonumber\\
&&-\xi_{td}(C_3+\frac{1}{3}C_4+C_9+\frac{1}{3}C_{10})+\xi_{td}(C_5+\frac{1}{3}C_6+C_7+\frac{1}{3}C_{8}) \Big]\nonumber\\
&&+F_{a, \pi \rho}\Big[-\xi_{td}(\frac{4}{3}C_3+\frac{4}{3}C_4-\frac{2}{3}C_9-\frac{2}{3}C_{10})+\xi_{td}(C_5+\frac{1}{3}C_6-\frac{1}{2}C_7-\frac{1}{6}C_{8}) \Big]\nonumber\\
&&+F^P_{a,\pi\rho }\Big[
-\xi_{td}(\frac{1}{3}C_5+C_6-\frac{1}{6}C_7-\frac{1}{2}C_8) \Big],
\end{eqnarray}

\begin{eqnarray}\label{Mp0r-}
\sqrt{2}\mathcal{M}(B^-\rightarrow\pi^0 \rho^-)=&&(F_{e,\pi \rho}+F_{a,\rho \pi}-F_{a,\pi \rho})\Big[ \xi_{ud}(\frac{1}{3}C_1+C_2)-\xi_{td}(\frac{1}{3}C_3+C_4+\frac{1}{3}C_9+C_{10}) \Big]\nonumber\\
&&+F_{e,\rho \pi }\Big[ \xi_{ud}(C_1+\frac{1}{3}C_2)-\xi_{td}(-\frac{1}{3}C_3-C_4+\frac{5}{3}C_9+C_{10}-\frac{3}{2}C_7-\frac{1}{2}C_8) \Big]\nonumber\\
&&+F^P_{e, \rho\pi}\Big[  \xi_{td}(\frac{1}{3}C_5+C_6-\frac{1}{6}C_7-\frac{1}{2}C_{8}) \Big]+(M_{e,\pi \rho}+M_{a,\rho \pi}-M_{a,\pi \rho})\Big[ \xi_{ud}(\frac{1}{3}C_1)\nonumber\\
&&-\xi_{td}(\frac{1}{3}C_3+\frac{1}{3}C_9) \Big]+M_{e, \rho\pi}\Big[ \xi_{ud}(\frac{1}{3}C_2)-\xi_{td}(-\frac{1}{3}C_3+\frac{1}{6}C_9+\frac{1}{2}C_{10}) \Big]\nonumber\\
&&+M^R_{e,\pi \rho}\Big[ -\xi_{td}(\frac{1}{3}C_{5}+\frac{1}{3}C_{7}) \Big]+M^P_{e, \rho\pi}\Big[ -\xi_{td}(\frac{1}{2}C_{8}) \Big]-M^R_{e, \rho\pi}\Big[ -\xi_{td}(\frac{1}{3}C_{5}-\frac{1}{6}C_{7}) \Big]\nonumber\\
&&+(M^R_{a, \rho\pi}-M^R_{a, \rho\pi})\Big[ -\xi_{td}(\frac{1}{3}C_{5}+\frac{1}{3}C_{7}) \Big] \nonumber\\
&&+(F^P_{a,\rho\pi }-F^P_{a,\pi\rho })\Big[-\xi_{td}(\frac{1}{3}C_5+C_6+\frac{1}{3}C_7+C_{8}) \Big]
\end{eqnarray}

\begin{eqnarray}\label{Mp-r0}
\sqrt{2}\mathcal{M}(B^-\rightarrow\pi^- \rho^0)=&&(F_{e, \rho\pi}+F_{a, \pi\rho}-F_{a, \rho\pi})\Big[ \xi_{ud}(\frac{1}{3}C_1+C_2)-\xi_{td}(\frac{1}{3}C_3+C_4+\frac{1}{3}C_9+C_{10}) \Big]\nonumber\\
&&+F_{e, \pi\rho }\Big[ \xi_{ud}(C_1+\frac{1}{3}C_2)-\xi_{td}(-\frac{1}{3}C_3-C_4+\frac{5}{3}C_9+C_{10}-\frac{3}{2}C_7-\frac{1}{2}C_8) \Big]\nonumber\\
&&+F^P_{e, \rho\pi}\Big[  -\xi_{td}(\frac{1}{3}C_5+C_6+\frac{1}{3}C_7+C_{8}) \Big]+(M_{e, \rho\pi}+M_{a, \pi\rho}-M_{a, \rho\pi})\Big[ \xi_{ud}(\frac{1}{3}C_1)\nonumber\\
&&-\xi_{td}(\frac{1}{3}C_3+\frac{1}{3}C_9) \Big]+M_{e, \pi\rho}\Big[ \xi_{ud}(\frac{1}{3}C_2)-\xi_{td}(-\frac{1}{3}C_3+\frac{1}{6}C_9+\frac{1}{2}C_{10}) \Big]\nonumber\\
&&+M^R_{e, \rho\pi}\Big[ -\xi_{td}(\frac{1}{3}C_{5}+\frac{1}{3}C_{7}) \Big]+M^P_{e, \pi\rho}\Big[ -\xi_{td}(\frac{1}{2}C_{8}) \Big]\nonumber\\
&&-M^R_{e, \pi\rho}\Big[ -\xi_{td}(\frac{1}{3}C_{5}-\frac{1}{6}C_{7}) \Big]+(M^R_{a, \pi\rho}-M^R_{a, \pi\rho})\Big[ -\xi_{td}(\frac{1}{3}C_{5}+\frac{1}{3}C_{7}) \Big] \nonumber\\
&&+(F^P_{a,\pi\rho }-F^P_{a,\rho\pi })\Big[-\xi_{td}(\frac{1}{3}C_5+C_6+\frac{1}{3}C_7+C_{8}) \Big]
\end{eqnarray}

\begin{eqnarray}\label{Mp0r0}
2\mathcal{M}(\bar{B^0}\rightarrow\pi^0 \rho^0)=&&(F_{e,\pi \rho}+F_{e, \rho\pi})\Big[ -\xi_{ud}(C_1+\frac{1}{3}C_2)-\xi_{td}(\frac{1}{3}C_3+C_4-\frac{5}{3}C_9 -C_{10}) \Big]-F^P_{e,\rho \pi} \xi_{td}\Big[ \frac{1}{3}C_5+C_6 \nonumber\\
&&-\frac{1}{6}C_7-\frac{1}{2}C_{8} \Big]-(F_{e,\pi \rho}-F_{e, \rho\pi})\Big[-\xi_{td}(\frac{3}{2}C_7+\frac{1}{2}C_{8})\Big]+(M_{e,\pi \rho}+M_{e,\rho\pi })\Big[ \nonumber\\
&&-\xi_{ud}(\frac{1}{3}C_2)-\xi_{td}(\frac{1}{3}C_3-\frac{1}{6}C_9-\frac{1}{2}C_{10}) \Big] +(M^P_{e,\pi \rho}+M^P_{e,\rho \pi})\Big[ -\xi_{td}(-\frac{1}{2}C_8) \Big] \nonumber\\
&&+(M_{a,\pi \rho}+M_{a,\rho \pi})\Big[ \xi_{ud}(\frac{1}{3}C_2)-\xi_{td}(\frac{1}{3}C_3+\frac{2}{3}C_4-\frac{1}{6}C_9  +\frac{1}{6}C_{10}) \Big]\nonumber
\end{eqnarray}
\begin{eqnarray}
&&+(M^R_{a,\pi \rho}+M^R_{a,\rho \pi}+M^R_{e,\pi \rho}+M^R_{e,\rho \pi})\Big[ -\xi_{td}(\frac{1}{3}C_5-\frac{1}{6}C_7) \Big]+(M^P_{a,\pi \rho}+M^P_{a,\rho \pi})\Big[ -\xi_{td}(\frac{2}{3}C_6+\frac{1}{6}C_{8}) \Big] \nonumber\\
&&(F_{a,\pi \rho}+F_{a,\rho \pi})\Big[ \xi_{ud}(C_1+\frac{1}{3}C_2)-\xi_{td}(\frac{7}{3}C_3+\frac{5}{3}C_4+\frac{1}{3}C_9-\frac{1}{3}C_{10})+\xi_{td}(2C_5+\frac{2}{3}C_6\nonumber\\
&&+\frac{1}{2}C_7+\frac{1}{6}C_{8}) \Big]+(F^P_{a,\pi \rho}+F^P_{a, \rho\pi})\Big[ -\xi_{td}(\frac{1}{3}C_5+C_6-\frac{1}{6}C_7-\frac{1}{2}C_8) \Big],
\end{eqnarray}

\begin{eqnarray}\label{Mp0omega}
2\mathcal{M} (\bar{B^0}\rightarrow \pi^0 \omega)=&&\xi_{ud}\Big[ (C_1+\frac{1}{3}C_2)(-F_{e,\pi \omega}+F_{e,\omega \pi}+F_{a,\pi \omega}+F_{a,\omega \pi}) +\frac{1}{3}C_2(-M_{e,\pi \omega}+M_{e,\omega \pi}+M_{a,\pi \omega}+M_{a,\omega \pi})\Big]\nonumber\\
&&-\xi_{td}\Big[  (-\frac{1}{3}C_3-C_4+\frac{5}{3}C_9+C_{10})(F_{e,\omega \pi} +F_{a,\omega \pi}+F_{a,\pi \omega})+ \frac{3}{2}( C_7 +\frac{1}{3}C_8)(-F_{e,\omega \pi} -F_{a,\pi \omega}-F_{a,\omega \pi}) \nonumber\\
&&  + (-\frac{1}{3}C_5-C_6+\frac{1}{6}C_7+\frac{1}{2}C_{8})   (F^P_{e,\omega \pi}+F^P_{a,\pi \omega}+F^P_{a,\omega \pi})+(-\frac{7}{3}C_3-\frac{5}{3}C_4-\frac{1}{3}C_9+\frac{1}{3}C_{10}-2 C_5\nonumber \\
&&-\frac{2}{3}C_6-\frac{1}{2}C_7-\frac{1}{6}C_{8})F_{e,\pi \omega} +(-\frac{1}{3}C_3+\frac{1}{6}C_9+\frac{1}{2}C_{10})(M_{e,\omega \pi}+M_{a,\omega\pi}+M_{a,\pi \omega})\nonumber \\
&&+(-\frac{1}{3}C_5+\frac{1}{6}C_7)(M^R_{e,\pi \omega}+M^R_{e,\omega\pi}+M^R_{a,\pi \omega}+M^R_{a,\omega\pi})+(\frac{1}{2}C_8)(M^P_{e,\omega\pi}+M^P_{a,\pi \omega}+M^P_{a,\omega \pi})\nonumber \\
&&+(-\frac{1}{3}C_3-\frac{2}{3}C_4+\frac{1}{2}C_9-\frac{1}{2}C_{10}) M_{e,\pi \omega}+(-\frac{2}{3}C_6-\frac{1}{6}C_8)M^P_{e,\pi \omega} \Big]
\end{eqnarray}

\begin{eqnarray}\label{Mp-omega}
\sqrt{2}\mathcal{M} (B^-\rightarrow \pi^- \omega)=&&\xi_{ud}\Big[ (\frac{1}{3}C_1+C_2)(F_{e,\omega \pi}+F_{a,\pi \omega}+F_{a,\omega \pi}) +(C_1+\frac{1}{3}C_2)F_{e,\pi \omega}+(\frac{1}{3}C_2)M_{e,\pi \omega}\nonumber \\
&&+(\frac{1}{3}C_1)(M_{e,\omega \pi}+M_{a,\pi \omega}+M_{a,\omega \pi})\Big]-\xi_{td}\Big[(\frac{7}{3}C_3+\frac{5}{3}C_4+\frac{1}{3}C_9-\frac{1}{3}C_{10}+2 C_5+\frac{2}{3}C_6+\frac{1}{2}C_7\nonumber \\
&&+\frac{1}{6}C_{8})F_{e,\pi \omega}  +(\frac{1}{3}C_3+C_4+\frac{1}{3}C_9+C_{10})F_{e,\omega \pi} + (\frac{1}{3}C_5+C_6+\frac{1}{3}C_7+C_{8})F^P_{e,\omega\pi}\nonumber \\
&& +(\frac{1}{3}C_3+\frac{2}{3}C_4-\frac{1}{6}C_9+\frac{1}{6}C_{10})M_{e,\pi \omega} +(\frac{1}{3}C_5-\frac{1}{6}C_7)M^R_{e,\pi \omega}+(\frac{2}{3}C_6+\frac{1}{6}C_8)M^P_{e,\pi \omega}\nonumber\\
&& +(\frac{1}{3}C_3+\frac{1}{3}C_9)M_{e,\omega \pi}+ (\frac{1}{3}C_5+\frac{1}{3}C_7)M^R_{e,\omega\pi}+(\frac{1}{3}C_3+\frac{1}{3}C_9)(M_{a,\pi \omega}+M_{a,\omega \pi}) \nonumber \\
&&+(\frac{1}{3}C_3+C_4+\frac{1}{3}C_9+C_{10})(F_{a,\pi \omega}+F_{a,\omega \pi}) + (\frac{1}{3}C_5+C_6+\frac{1}{3}C_7+C_{8})(F^P_{a,\pi \omega}+F^P_{a,\omega \pi})\nonumber \\
&&+(\frac{1}{3}C_3+\frac{1}{3}C_9)(M_{a,\pi \omega}+M_{a,\omega \pi}) + (\frac{1}{3}C_5+\frac{1}{3}C_7)(
M^R_{a,\pi \omega}+M^R_{a,\omega \pi})  \Big],
\end{eqnarray}

\end{widetext}
where $\xi_{ud} = V_{ub}V_{ud}^{*}$, $\xi_{td}=V_{tb} V_{td}^{*}.$  The decay width and $CP$ violation are expressed as
\begin{equation}
\Gamma(B \rightarrow f) =\frac{G_F^2 m_B^3}{128 \pi} |\mathcal{M}(B\rightarrow f)|^2,
\end{equation}
and
\begin{equation}
A_{CP}(B \rightarrow f) =\frac{\Gamma(\bar{B} \rightarrow \bar{f})- \Gamma(B \rightarrow f) }{\Gamma(\bar{B} \rightarrow \bar{f})+ \Gamma(B \rightarrow f)}.
\end{equation}

\section{Next-to-Leading Order Corrections to the Hard Amplitude}

To improve the accuracy of the calculations, we will use Wilson coefficients corrected to next-to-leading order (NLO) in QCD and consider the most significant contributions at NLO. These include vertex corrections, quark loops, and magnetic penguin contributions. The NLO contributions can be attributed to the modifications of the Wilson coefficients. For convenience, we define the combination of Wilson coefficients as follows
\begin{eqnarray}
&&	a_1(\mu) =C_2(\mu) +\frac{C_1(\mu)}{N_c}, \nonumber\\
&&	a_2(\mu) =C_1(\mu) +\frac{C_2(\mu)}{N_c}, \nonumber\\
&&	a_i(\mu) =C_i(\mu) +\frac{C_{i\pm 1}(\mu)}{N_c}, ~~ i=3-10  
\end{eqnarray}
In the above expression, the plus (minus) sign is taken when $i$ is odd (even).

\subsubsection{Vertex Correction}

The vertex corrections modify the Wilson coefficients as follows \cite{QCDf1,QCDf2,QCDf3,LiMiSa2005}
\begin{eqnarray}
&&	a_1(\mu) \rightarrow a_1(\mu)+ \frac{\alpha_s(\mu)}{4 \pi}C_F\frac{C_1(\mu)}{N_c}V_1 ,\nonumber\\
&&	a_2(\mu) \rightarrow a_2(\mu)+ \frac{\alpha_s(\mu)}{4 \pi}C_F\frac{C_2(\mu)}{N_c}V_2 ,\\
&&	a_i(\mu) \rightarrow a_i(\mu)+ \frac{\alpha_s(\mu)}{4 \pi}C_F\frac{C_{i \pm 1}(\mu)}{N_c}V_i ,\space   i=3-10, \nonumber
\end{eqnarray}

In the naive dimensional regularization (NDR) scheme, the function $V_i$ for the pseudoscalar meson is expressed as \cite{QCDf1,QCDf2,QCDf3}:

\begin{widetext}
	
	\begin{equation}
	V_i=\left\{
	\begin{aligned}
	&12\ln\frac{m_b}{\mu}-18+\int_{0}^{1}dx \phi_{\pi}(x)g(x),& \mathrm{for} \; i =1-4,9,10, \\
	-&12\ln\frac{m_b}{\mu} + 6 -\int_{0}^{1}dx \phi_{\pi}(x)g(1-x),& \mathrm{for} \; i =5,7,   \\
	-&6 +\int_{0}^{1}dx \phi^P_{\pi}(x)h(1-x),& \mathrm{for} \; i= 6,8
	\end{aligned}
	\right.
	\end{equation}
	where $\phi_{\pi}(x)$ and $\phi^P_{\pi}(x)$ are the distribution amplitudes of twist-2 and -3 for the emitted meson, respectively. For vector mesons, the wave functions $\phi_{\pi}(x)$ and $\phi^P_{\pi}(x)$  need to be replaced with $\phi_{\rho,\omega}(x)$ and $(-\phi^s_{\rho,\omega}(x))$, respectively \cite{LiHN-SM2006}. The hard kernels $g(x)$ and $h(x)$ are
	\begin{equation}
	g(x) = 3 \left( \frac{1-2x}{1-x} \ln x -i \pi \right)+ \left[ 2\mathrm{Li}_2(x)- \ln^2 x-\frac{2 \ln x}{1-x} -(3+2i\pi)\ln x - (x \leftrightarrow 1 -x)  \right],
	\end{equation}
	\begin{equation}
	h(x) =   2\mathrm{Li}_2(x)- \ln^2 x -(1+2i\pi)\ln x - (x \leftrightarrow 1 -x)  .
	\end{equation}
	
\end{widetext}


\subsubsection{The Quark-Loop Contributions}

For the $b\rightarrow d$ transition, the effective Hamiltonian for the quark-loop correction is given by \cite{LiMiSa2005}

\begin{eqnarray}
	\mathcal{H}_{\mathrm{eff}} &= &-\sum_{q=u,c,t}\sum_{q'}\frac{G_F}{\sqrt{2}}V_{qb}V_{qd}^*\frac{\alpha_s(\mu)}{2\pi}C^{(q)}(\mu,l^2)  \nonumber \\
	&&\times (\bar{d}\gamma_{\rho}(1-\gamma_5)T^a b)(\bar{q}'\gamma^{\rho}T^a q'),
\end{eqnarray}
where $l^2$ is the momentum squared of the virtual gluon connecting the quark loop and the final quark-antiquark pair. For $q = u, c$, the function $C^{(q)}(\mu,l^2)$ is
\begin{equation}\label{cq}
		C^{(q)}(\mu,l^2) =\left[G^{(q)}(\mu,l^2) -\frac{2}{3} \right]C_2(\mu).
\end{equation}
For $q = t$, the function $C^{(q)}(\mu,l^2)$ is
\begin{eqnarray}\label{ct}
	C^{(t)}(\mu,l^2) &=&\left[  G^{(d)}(\mu,l^2) -\frac{2}{3}\right]C_3(\mu) \nonumber \\
		&+&  \sum_{q''=u,d,s,c}G^{(q'')}(\mu,l^2)\left[C_4(\mu) -C_6(\mu) \right]. \nonumber \\
\end{eqnarray}
The function $G$ in Eqs. (\ref{cq}) and (\ref{ct}) is
\begin{equation}
	 G^{(q)}(\mu,l^2) = -4 \int_{0}^{1}dx x(1-x) \ln \frac{m_q^2-x(1-x)l^2-i\varepsilon}{\mu^2},
\end{equation}
where $m_q$ is the quark mass for $q=u,d,s,c$.

Since the effective Hamiltonian for the quark-loop correction shares the same topological structure as the penguin operators $O_4$ to $O_7$, its contribution can be absorbed into the combinations of Wilson coefficients $a_4$ and $a_6$ as
\begin{equation} \label{l-square}
a_{4,6}(\mu) \rightarrow a_{4,6}(\mu)+\frac{\alpha_s(\mu)}{9\pi}\sum_{q=u,c,t}\frac{V_{qb}V_{	qd}^*}{V_{tb}V_{td}^*}C^{(q)}(\mu,\left<l^2\right>).
\end{equation}
Here, $\langle l^2 \rangle = m_b^2 / 4$ is taken as the reasonable mean value of the momentum squared of the virtual gluon connecting the quark loop for $B$ decays.

\subsubsection{Magnetic Penguins}

The effective Hamiltonian of magnetic penguin for the weak $ b \rightarrow d\textsl{g} $ transition is
\begin{equation}\label{H-mpg}
	\mathcal{H}_{\mathrm{eff}}= - \frac{G_F}{\sqrt{2}}V_{tb}V_{td}^*C_{8\textsl{g}}O_{8\textsl{g}},
\end{equation}
where the magnetic-penguin operator is
\begin{equation}
	O_{8\textsl{g}}=\frac{g}{8 \pi^2}m_b \bar{d}_i \sigma_{\mu \nu}(1+\gamma_5)T_{ij}^a G^{a\mu\nu} b_j.
\end{equation}
The contribution of the Hamiltonian in Eq. (\ref{H-mpg}) can be absorbed into the relevant Wilson coefficients \cite{LiMiSa2005}
\begin{equation}
	a_{4,6}(\mu) \rightarrow a_{4,6}(\mu)-\frac{\alpha_s(\mu)}{9\pi} \frac{2m_B}{\sqrt{\left<l^2\right>}}C_{8\textsl{g}}^{\mathrm{eff}}(\mu),
\end{equation}
where the effective coefficient $C_{8\textsl{g}}^{\mathrm{eff}}=C_{8\textsl{g}}+C_5$ \cite{Hamiltanion1996}.

\vspace{0.5em}

\section{ Soft Form Factors and Color-Octet Contribution}

\subsection{ The contribution of soft form factors}
In the numerical analysis, we find that the contributions from diagrams (a), (b), (g), and (h) in Fig.\ref{fig1} still contain significant soft contributions in the perturbative calcuations \cite{Lu-Yang2021,lu-yang2023,wang-yang2023}. The reason originates from the different asymptotic behavior in the longitudinal endpoint region of the $B$-meson wave function derived from the QCD-inspired relativistic potential model, which weakened the suppression effect to the soft contributions. To maintain the reliability of perturbative calculations, a cutoff scale $\mu_c$ must be introduced. Perturbative calculations are applied only for contributions above the critical cutoff scale $\mu_c$. Meanwhile, for contributions below the cutoff scale, soft form factors are introduced.

Fig.\ref{fig1} (a) and (b) are related to the $B \to M$ transition form factor. The total transition form factor can be divided into two parts:
\begin{equation} \label{softBM}
F_+^{BM}=h_+^{BM}+\xi^{BM},
\end{equation}
where $h_+^{BM}$ is the hard $B \to M$ transition form factor, which can be calculated using perturbative QCD methods, and $\xi^{BM}$ is the soft transition form factor. The total amplitude is modifieded by the contribution of soft transition form factor as follows

\begin{eqnarray}
\mathcal{M} \rightarrow \mathcal{M}&-&2if_{\pi}C(\mu_{c})V_{\mathrm{CKM}} \cdot\xi^{BM}\nonumber\\
&-&4i\frac{\mu_M}{m_B}f_{M}C'(\mu_{c})V_{\mathrm{CKM}} \cdot\xi^{BM},
\end{eqnarray}
where $C(\mu_c)$ and $C'(\mu_c)$ are appropriate combinations of the Wilson coefficients corresponding to $(V-A)(V-A)$ and $(S+P)(S-P)$ current operators, respectively, which are evaluated at the cutoff scale $\mu$.

The soft contributions from diagrams (g) and (h) can be described by the soft production form factor of $M_1M_2$. The production form factor can be defined by the meson matrix element. However, for the $B \to \pi\rho$ process,
\begin{equation}\label{PV-formfactor}
\langle \pi \rho|S-P|0\rangle =-\langle \rho \pi|S-P|0\rangle,
\end{equation}
where the order of the mesons follows the convention that the first particle is $M_1$ and the second is $M_2$. The total production form factor in $B \to \pi\rho$ decays can be defined as
\begin{equation}\label{s-formfactor}
\langle \rho \pi |S-P|0\rangle =-\dfrac{1}{2}\sqrt{\mu_{\pi}m_{\rho}} F_{+}^{\rho \pi}(q^2).
\end{equation}
Here $\mu_{\pi}$ is the chiral mass for pion, and $m_\rho$ is the mass of the $\rho$ meson. The total production form factor can also be divided into hard and soft parts
\begin{equation} \label{s-form-hard-and-soft}
F_{+}^{\rho \pi}=h^{\rho \pi}+\xi^{\rho \pi}.
\end{equation}
The production form factor in $B \to \pi\omega$ decays is similar to this.

The soft production form factor will be treated as a phenomenological input parameter, and its contribution to the decay amplitude shows up in the following way
\begin{equation}
\mathcal{M} \rightarrow \mathcal{M}+\frac{2\sqrt{\mu_{\pi}m_{\rho(\omega)}}}{m_B^2} \langle0|S-P|B\rangle C(\mu_{c})V_{\mathrm{CKM}} \xi^{M_1M_2},
\end{equation}
where $\langle0|S-P|B\rangle=-i\chi_B$, and $\chi_B$ can be found in Eq. (\ref{chiB}).

\subsection{The color-octet contribution}

Since the final-state mesons observed in experiments are color singlets, one usually only considers the contribution of quark-antiquark pair of color-singlet state in theoretical calculations in treating nonleptonic two-body $B$ decays. The contribution of quark-antiquark pairs in color-octet state are generally ignored. However, color-octet quark-antiquark pairs can appear as intermediate states in $B$-meson decays and subsequently transform into color-singlet sates through exchanging soft gluons in long-distance hadronic scale. Thus, the production of color-octet state in hard interaction can contribute to the final decay amplitude. This mechanism has been considered in the calculations of many rare $B$-meson decay processes by us recently \cite{lu-yang2023,wang-yang2023}, which can simultaneously resolve the $\pi\pi$ and $K\pi$ puzzles and provid more accurate predictions for branching ratios and $CP$ violations for a number of experimentally unmeasured decay modes. Here we will continue to employ this mechanism to study the $B \to \pi\rho(\omega)$ decays.

\begin{figure}[htb]
	\includegraphics[width=0.40\textwidth]{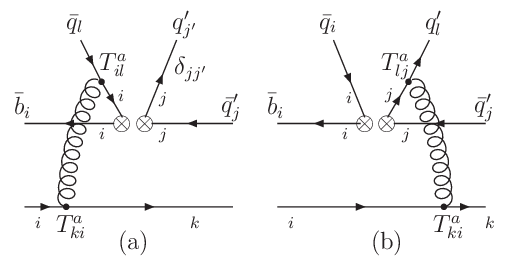}
	\caption{\label{fig-color8} Two nonfactorizable diagrams with operator insertion of $(\bar{b}_iq_i)(\bar{q}^\prime_j q^\prime_j)$, where the explicit type of the current is omitted. The quark-antiquark pairs in the final state are in color non-singlet states. The symbols $i$, $j$, $j^\prime$, $k$ and $l$ are color indices.}
\end{figure}

The contributions from color-octet states can be identified by analyzing the structure of color factors. The main formula used is the identity relation for the generators of the color SU(3) group
\begin{equation}\label{COIR}
T^{a}_{ij}T^{a}_{kl}=-\dfrac{1}{2N_{c}}\delta_{ij}\delta_{kl}+\dfrac{1}{2}\delta_{il}\delta_{kj}.
\end{equation}
Considering the quark-antiquark pairs in the diagrams in Fig. 1 as color-octet state and using Eq.(\ref{COIR}), we can derive the color-octet contributions to the amplitude. For example, for the topology of diagrams (c) and (d) in Fig. 1, we illustrate the Feynman diagrams with final quark-antiquark pairs in the color-octet state in Fig. \ref{fig-color8}.

The summation of color factors for Fig. \ref{fig-color8} (a) yields

\begin{equation}\label{eq:cfa}
\begin{split}
\sum_{ijkl} T_{ki}^a T_{il}^a &= \sum_{jkl} C_F \delta_{lk}
= \sum_{jklj^\prime} C_F \delta_{lk} \delta_{jj^\prime} \\
& = \sum_{jklj^\prime} C_F \left(\dfrac{1}{N_c} \delta_{lj^\prime} \delta_{jk} + 2 T_{lj^\prime}^a T_{jk}^a\right),
\end{split}
\end{equation}
The first term in the above expression corresponds to the color-singlet contribution, while the second term corresponds to the color-octet contribution. The result for Fig.~\ref{fig-color8} (b) is
\begin{equation}\label{eq:cfb}
\begin{split}
&\sum_{ijkl} T_{lj}^a T_{ki}^a
= \sum_{ijkl} \left[-\dfrac{1}{2N_c} \delta_{lj} \delta_{ki} + \dfrac{1}{2} \delta_{li} \delta_{kj} \right] \\
&\quad = \sum_{ijkl} \left[-\dfrac{1}{2N_c} \left(\dfrac{1}{N_c} \delta_{li} \delta_{kj}
+ 2 T_{li}^b T_{kj}^b\right) + \dfrac{1}{2} \delta_{li} \delta_{kj} \right]\\
&\quad = \sum_{ijkl} \left(\dfrac{C_F}{N_c} \delta_{li} \delta_{kj}
- \dfrac{1}{N_c} T_{li}^b T_{kj}^b \right),
\end{split}
\end{equation}
The first terms in Eqs.~(\ref{eq:cfa}) and ~(\ref{eq:cfb}), representing the color-singlet contributions, which give $M_e$, $M_e^R$, and $M_e^P$. By analyzing the relationship between the second term and the first term, we obtain the amplitude for the contribution of theintermediate color-octet state
\begin{equation}\label{eq:fe8}
\mathcal{M}_e^{(P,R)8} \equiv 2N_c^2 \mathcal{M}_e^{(P,R)c} - \dfrac{N_c}{C_F} \mathcal{M}_e^{(P,R)d}.
\end{equation}
Here, it is assumed that the parton distribution functions of the color-octet states are the same as thatof color-singlet meson state.

The color-octet states shoulbe be changed into color-singlet mesons by exchanging soft gluonse, which is essentially of non-perturbative dynamics. We can describe it by introducing a few phenomenological parameters. Thus, the final contribution is
\begin{equation}\label{nfMe8}
Y^8_M \mathcal{M}_e^{(P,R)8},
\end{equation}
The subscript $M$ indicates that the contribution is from non-factorizable diagram, while the superscripts $P$ and $R$ denote the insertion of operators of $(S+P)(S-P)$ and $(V-A)(V+A)$ current, respectively. The absence of a superscript indicates the insertion of a $(V-A)(V-A)$ current operator.

The non-zero contributions from the other diagrams in Fig. 1 are:
\begin{equation}
Y^8_F F_e^{(P)8},\;
Y^8_M\mathcal{M}_e^{(P,R)\prime 8},\;Y^8_M\mathcal{M}_a^{(P,R)8},\; Y^8_F F_a^{(P)8},
\end{equation}
where
\begin{equation}\label{fa8}
\begin{split}
& F_e^{(P)8} \equiv 2N_c^2 F_e^{(P)a} - \dfrac{N_c}{C_F} F_e^{(P)b}, \\
&\mathcal{M}_e^{(P,R)\prime 8} \equiv \dfrac{N_c^2}{C_F} \mathcal{M}_e^{(P,R)}, \
\mathcal{M}_a^{(P,R)8} \equiv -\dfrac{N_c}{C_F} \mathcal{M}_a^{(P,R)}\\
&      F_a^{(P)8} \equiv -\dfrac{N_c^2}{C_F} F_a^{(P)},
\end{split}
\end{equation}

The quantities $F_e^{(P)a}$, $F_e^{(P)b}$, $\mathcal{M}_e^{(P,R)c}$, $\mathcal{M}_e^{(P,R)d}$,  $\mathcal{M}_a^{(P,R)}$ and $F_a^{(P)}$ are the convolution functions corresponding to the diagrams (a) $\sim$ (h) in Fig.~\ref{fig1} by using the PQCD approach.

\section{Numerical analysis and discussion}

In numerical calculations, we require some non-perturbative phenomenological input parameterss. These inputs, in addition to the parameters in the meson wave functions (provided in the Appendix B), include meson masses, decay constants, soft form factors and color-octet parameters. The meson masses and decay constants are \cite{PDG2024,SY2019}
\begin{equation}\label{Const}
\begin{split}
& f_B = 0.210\;\mathrm{GeV},   \;\;\;\;  m_B=5.2792 \;\mathrm{GeV}\\
& f_{\pi} = 0.130\;\mathrm{GeV},\;\;\;\;\;  \mu_{\pi} = 1.75\;\mathrm{GeV}\\
&f_{\rho}^{||} = 0.216\; \mathrm{GeV},   \;\;\;\;   f_{\rho}^{\perp} = 0.165\; \mathrm{GeV}\\
& f_{\omega}^{||} = 0.195\; \mathrm{GeV},  \;\; \;\;  f_{\omega}^{\perp}=0.145\;\mathrm{GeV},\\
& m_{\rho} = 0.775\;\mathrm{GeV}, \;\;\;\;     m_{\omega} = 0.782\;\mathrm{GeV},
\end{split}
\end{equation}

By considering non-perturbative methods such as light-cone sum rules and experimental measurements of $B$-meson semileptonic decays \cite{Belle2013,ball2005new,PBall2005BV,RZwicky2016,LuWang2020,MAI2007,MAIS2012}, we obtain the total physical transition form factors as
\begin{eqnarray}
&&F_+^{B\pi} = 0.27 \pm0.02,  \nonumber \\
&&A_0^{B \rho} = 0.32 \pm0.05,    \\
&&A_0^{B \omega} = 0.32 \pm 0.05. \nonumber
\end{eqnarray}
For the $B\to\pi$ transition form factor, $F_+^{B\pi}$ is derived from the measured data of semileptonic $B\to\pi$ decays \cite{Belle2013}. The numerical values of $A_0^{B \rho}$ and $A_0^{B \omega}$ are taken as the averaged values from Light-Cone Sum rules calculations \cite{PBall2005BV,RZwicky2016}, which agrees with results from soft-collinear effective theory and quark model predictions \cite{LuWang2020,MAI2007,MAIS2012}.

For contributions with energy scale large than the cutoff scale $\mu>\mu_c $, where $\mu_c = 1 \, \text{GeV}$ is taken, we use PQCD approach to calculate the hard transition form factors. The results are
\begin{eqnarray}
&&h_+^{B\pi} = 0.23 \pm0.01,  \nonumber \\
&&h_{A_0}^{B \rho } = 0.18 \pm0.01,    \\
&&h_{A_0}^{B\omega} = 0.18 \pm0.01.  \nonumber
\end{eqnarray}
Then, according to Eq.~(\ref{softBM}), the soft $B\to M$ transition form factor can be obtained as
\begin{eqnarray}
&&\xi^{B\pi} = 0.04 \pm0.01,  \nonumber \\
&&\xi^{B \rho}_{A_0} = 0.14 \pm0.04,    \\
&&\xi^{B\omega}_{A_0} = 0.14 \pm0.04 . \nonumber
\end{eqnarray}

For the soft production form factor $\xi$ and the color-octet factors $Y_F^8$ and $Y_M^8$, they are parameters describing contributions of non-perturbative long-distance interactions. Their values can not be calculated reliablely in theory at present, which can only be determined by fitting experimental data currently. We employ the chi-square analysis method to estimate these parameters. The definition of chi-square statistic is given by

\begin{equation}
\chi^2 = \sum_{i=1}^N \frac{(y_i - \mu_i(\hat\theta))^2}{\sigma_i^2},
\end{equation}
where $y_i$ represents the experimentally measured values (such as decay branching ratios and direct $CP$-violation values), $\mu_i$ denotes the relevant physical quantities calculated with the selected parameter set $\hat\theta=(\theta_1,\cdots,\theta_M)$ ($M$ is a non-zero integer), which can be the soft input parameters in this work, and $\sigma_i$ represents the standard deviations of the experimental data.

A critical metric in this framework is the reduced chi-square statistic $\chi^2_{\mathrm{red}}$, which normalizes the chi-square value by the degrees of freedom (DoF).
\begin{equation}
\chi^2_{\text{red}} = \frac{\chi^2}{N - k},
\end{equation}
where $N$ represents the number of data points, $k$ is the number of parameters, and $N - k$ denotes the degrees of freedom.

For simplicity and to decrease the number of unknown input parameters in this work, we assume that the soft parameters for $\rho\pi$ and $\omega\pi$ chunnels are the same. Consider all the measured branching ratios and $CP$ violations for  $B\to\rho\pi$ and $\omega\pi$ decays, we finally obtain the values for the input parameters as follows
\begin{eqnarray}
&&Y^8_F=(0.097 \pm 0.019) e^{(-0.743 \pm 0.028)\pi i},  \nonumber \\
&&Y^8_M=(0.0722\pm 0.0031) e^{(0.936\pm0.037)\pi i},    \\
&&\xi=(0.286 \pm 0.026)e^{(-0.387\pm 0.041)\pi i}. \nonumber
\end{eqnarray}
The corresponding reduced chi-square statistic is:
\begin{equation}
\chi ^2_{\mathrm{red}} = \frac{5.9}{11-6}=1.18,
\end{equation}

The comparison between theoretical calculations and experimental measurements of decay branching ratios and direct $CP$ violations are presented in Table \ref{table-BRCP-I1}.

\newcolumntype{H}{>{\setbox0=\hbox\bgroup}c<{\egroup}@{}}		
\begin{table*}[htb]
	\renewcommand\arraystretch{1.5}
	\caption{\label{table-BRCP-I1} Branching ratio ( $\times 10^{-6}$) and direct $CP$ violation with NLO contributions for decay modes of $\pi\rho$, $\pi \omega$  final states. Column ``NLO$^*$" incorporates additional contributions from soft form factors. NLO+soft includes all soft contributions. Column ``Data" is for the averaged values given in PDG \cite{PDG2024}. Experimental values of Data$^*$ are taken from the data of BABAR experimental collaboration \cite{BaBar2009L,BaBar2003T,BaBar2007AA}. }
	\begin{threeparttable}
		\setlength{\tabcolsep}{2.5mm}{
			\begin{tabular}{c|c|c|HHHcHHH|c|c|c}
				\hline
				\hline
				&$\mathrm{LO_{NLOWC}}$		&  NLO  &NLO*&$\mathrm{LO^V_{NLOWC}}$&$\mathrm{LO^*_{NLOWC}}$	& NLO$^*$  & NLO+soft & NLO& NLO+soft* & NLO+soft   & Data$^*$	& Data \cite{PDG2024}   \\ \hline
				Br($B^0\rightarrow \pi^\pm \rho^\mp$) &15.7	& 17.7  &21.6 &23.7 &29.6& 33.5 & 23.4 &  23.5 & 24.3  & $24.7\pm1.1^{+1.1+0.3}_{-1.3-0.3}$ &  	 	&$23.0\pm2.3$   \\
				
				Br($B^+\rightarrow \pi^0 \rho^+$)    &6.70	& 7.08 & 9.87 &5.32&8.19 & 13.1 & 13.2 & 10.2 & 13.1   & $12.6\pm0.9^{+0.6+0.32}_{-0.7-0.33}$  &     &   $10.6^{+1.2}_{-1.3}$     \\
				
				Br($B^+\rightarrow \pi^+ \rho^0$) &4.18	& 3.05 & 3.68 & 9.55 &8.81	& 4.25& 5.74 & 7.7& 5.73  & $6.04\pm0.47^{+0.19+0.03}_{-0.24-0.02}$  & $8.1\pm1.7$\cite{BaBar2009L} &$8.3\pm1.2$   \\
				
				Br($B^0\rightarrow \pi^0 \rho^0$)& 0.23	&	0.03 &0.17 &0.46 & 0.38& 0.04 & 2.1 & 0.89 & 1.99  & $1.95\pm0.34^{+0.06+0.05}_{-0.05-0.05}$  &   & $2.0\pm0.5$   \\
				
				Br($B^0\rightarrow \pi^0 \omega$)    &0.02	& 0.005 & 0.06 & 0.09 &0.16 & 0.11 & 0.30 & 0.21 & 0.26  & $0.30\pm0.04^{+0.06+0.01}_{-0.07-0.01}$  &    &   $<0.5$     \\
				
				Br($B^+\rightarrow \pi^+ \omega$) &2.37	& 2.67 & 2.21 & 6.83  &6.53	& 7.08 & 6.9  & 7.4 & 6.90    & $7.17\pm0.57^{+0.32+0.09}_{-0.30-0.11}$ & 	  & $6.9\pm0.5$   \\
				\hline
				\hline
				$A_{CP}$($B^0\rightarrow \pi^- \rho^+$) &0.13	 & 0.06  & 0.06 &  0.13&0.09 & -0.19 & -0.04 & 0.03 &-0.12 &$-0.142\pm0.032^{+0.002+0.004}_{-0.003-0.004}$  & $-0.18 \pm0.09$\cite{BaBar2003T}  	&$0.13\pm0.06$    \\				
				$A_{CP}$($B^0\rightarrow \pi^+ \rho^-$)   & -0.36  &-0.37  & -0.36  & -0.29 & -0.29	&0.03 & -0.23  & -0.33  & -0.07  &	$-0.04\pm0.04^{+0.00+0.01}_{-0.00-0.01}$  &  	 & $-0.08\pm0.08$   \\
				
				$A_{CP}$($B^+\rightarrow \pi^0 \rho^+$) & 0.26 	& 0.20 & 0.20 & 0.37 & 0.30 & -0.15 & 0.04 & 0.23 & 0.015  & $0.017\pm0.051^{+0.004+0.003}_{-0.003-0.003}$  &  	 &$0.03\pm0.10$  \\
				
				$A_{CP}$($B^+\rightarrow \pi^+ \rho^0$) & -0.45 	& -0.38  & -0.39  & 0.37 & -0.33 & 0.25	& -0.008  & -0.30& -0.001  & $-0.0025\pm0.075^{+0.0045+0.0071}_{-0.0051-0.0059}$   & 	 &$0.003\pm0.014$  \\
				
				$A_{CP}$($B^0\rightarrow \pi^0 \rho^0$) & 0.06  &	0.84 & 0.23  &-0.06 & 0.05& 0.89 &  0.17 &  0.12& 0.25  & $0.23\pm0.05^{+0.01+0.01}_{-0.01-0.01}$  &$0.10 \pm 0.66$\cite{BaBar2007AA}  &	$-0.27\pm0.24$    \\
				
				$A_{CP}$($B^0\rightarrow \pi^0 \omega$)   & 0.80 & 0.87 & 0.40  & 0.69 & 0.47	& 0.46 & 0.38   &  0.49 &0.39 &$0.41\pm0.04^{+0.07+0.05}_{-0.07-0.04}$  &  & $...$   \\
				
				$A_{CP}$($B^+\rightarrow \pi^+ \omega$) & -0.02 & -0.34 & 0.02 & -0.08 & -0.06 &-0.26		&-0.04  & -0.12  & -0.05 &$-0.031\pm0.044^{+0.015+0.017}_{-0.018-0.013}$ &  & $-0.04\pm 0.05$  \\
				
				\hline
				\hline
		\end{tabular}  }		
	\end{threeparttable}
	
\end{table*}

The column ``$\mathrm{LO_{NLOWC}}$" shows the leading-order QCD contributions with NLO Wilson coefficients. The column ``NLO" includes the most significant NLO contributions. Column ``NLO$^*$" incorporates additional contributions from both the soft $B \to M$ transition form factor and the soft $M_1M_2$ production form factor. Column ``NLO+soft" represents the results that incorporate both the main NLO contributions and the long-distance effects from soft form factors and color-octet states. The uncertainties in the first column originate from the soft form factors and color-octet parameters, while the errors in the second and third columns stem from the wave function parameters of the $B$ meson and light mesons. The data in the last column are PDG averages \cite{PDG2024}, while column ``Data$^*$" presents results from BABAR experimental collaboration \cite{BaBar2009L,BaBar2003T,BaBar2007AA}.

The comparison of the the column NLO and $\mathrm{LO_{NLOWC}}$ indicates that the effect of the diagrams of next-to-leading order in QCD is really small, which shows the modified PQCD approach indeed improve the behavior of perturbative calculation. The difference of the numerical results in columns of NLO$^*$ and NLO gives the contribution of the soft form factors, which shows that the effects of the soft form factors in $B\to \rho\pi$ and $\omega\pi$ are large. The comparison of the columns NLO+soft and NLO$^*$ indicates the effect of the color-octet contributions in these decays, which shows that the color-octet contributions are significant for making the theoretical calculation be consistent with experimental data. Table \ref{table-BRCP-I1} shows that our results for both branching ratios and $CP$ violations are consistent with experimental data in general.


Similar as the case of $B\to\pi\pi$ decays, the experimentally measured branching fraction for $B$ meson decays into $\rho^0\pi^0$ significantly exceeds the earlier and recently unpgraded theoretical predictions from PQCD calculations \cite{LuYang2002,RGLu2012,WangLu2008,xiao2022}. Our theoretical approach substantially alleviates this discrepancy.

For direct $CP$ violation measurements, the most notable case is $A_{CP}(B^0\to\pi^0\rho^0)$. Both our calculations and previous PQCD results for this channel yield opposite signs compared to the PDG average, while QCDF provides predictions with large uncertainty \cite{QCDf4,cheng-chua2009b} and SCET gives results near zero \cite{WangLu2008}. Our present result agrees with the data from  \text{BABAR} collaboration. Similar situation occurs for $\mathcal{B}(B^0\to\pi^-\rho^+)$. We hope that future \text{Belle II} and upgraded \text{LHCb} experiments with accumulated statistics will present more confirmative measurements for these quantities.

Within the framework that we used to reslove the $\pi\pi$ and $K\pi$ puzzles previouly, we calculate branching fractions and $CP$ violations for $B\to\pi\rho(\omega)$ decays which give results in well agreement with present experimental data.  Notably, for $B^0\to\pi^0\omega$ decays, QCDF and SCET predict branching fractions of $\mathcal{O}(10^{-8})$ \cite{QCDf4,cheng-chua2009b,WangLu2008}, while our improved PQCD approach predicts values one order of magnitude larger than them. This difference can be tested by the upgraded \text{Belle II} and \text{LHCb} experiments.

\section{Summary}

We have re-examined the $B\to\pi\rho$ and $B\to\pi\omega$ decay processes using the modified PQCD approach. In this study, we employ a $B$-meson wave function obtained by solving heavy-light quark-antiqurk wave equation in the relativistic potential model, which has stronger dynamical foundations. By introducing a critical cutoff scale $\mu_c$, we achieve a clear separation between the perturbative contributions and non-perturbative long-distance effects, where the soft form factors are used to characterize contributions below the cutoff scale. This ensures the validity of the PQCD calculation. Additionally, we include contributions from color-octet intermediate states which are generated after the short-distance hard interaction and finally transfered into final-state mesons by exchanging soft gluons. By using statistical methods, we find a reasonable phenomenological parameter space that can generate theoretical results  consistent with experimental data on $B\to\pi\rho$ and $B\to\pi\omega$ decays, and predict branching ratio and $CP$ violation for $B^0\to\pi^0\omega$ decay, which have not been successfully measured in experiment yet.

\vspace{0.5cm}
\acknowledgments
This work is supported in part by the National Natural Science Foundation of China under
Contracts No. 12275139, 11875168.

\appendix{}
\section{\label{a}Sudakov factor and ultraviolet logarithms in QCD}

The threshold factor $S_t(x)$ can be parameterized as \cite{lihn2002,TK-HNL}
\begin{equation}
  S_t(x)=\frac{2^{1+2c}\Gamma(3/2+c)}{\sqrt{\pi}\Gamma(1+c)}[x(1-x)]^c,
\end{equation}
with $c=0.3$.

The exponentials $\exp[-S_{B}(\mu)]$,  $\exp[-S_{M_1}(\mu)]$ and $\exp[-S_{M_2}(\mu)]$ are the Sudakov factor and the relevant single ultraviolet logarithms associated with the heavy and light mesons. The exponents are \begin{equation}
S_B(\mu) = s(x,b,m_B)-\frac{1}{\beta_1}\ln \frac{\ln (\mu/\Lambda_{\mbox{QCD}})}
         {\ln (1/(b\Lambda_{\mbox{QCD}}))}
\end{equation}
\begin{eqnarray}
S_{M_1}(\mu) &=& s(x_1,b_1,m_B)+s(1-x_1,b_1,m_B)\nonumber\\
&&\;\;-\frac{1}{\beta_1}\ln \frac{\ln (\mu/\Lambda_{\mbox{QCD}})}
         {\ln (1/(b_1\Lambda_{\mbox{QCD}}))}
\end{eqnarray}
\begin{eqnarray}
S_{M_2}(\mu) &=& s(x_2,b_2,m_B)+s(1-x_2,b_2,m_B)\nonumber\\
&&\;\;-\frac{1}{\beta_1}\ln \frac{\ln (\mu/\Lambda_{\mbox{QCD}})}
         {\ln (1/(b_2\Lambda_{\mbox{QCD}}))}
\end{eqnarray}
\vspace{0.6cm}

The exponent $S(x,b,Q)$ up to next-leading order in QCD is \cite{Li1995}
\begin{widetext}
\begin{eqnarray}
&& s(x,b,Q)=\frac{A^{(1)}}{2\beta_{1}}\hat{q}\ln\left(\frac{\hat{q}}
{\hat{b}}\right)-
\frac{A^{(1)}}{2\beta_{1}}\left(\hat{q}-\hat{b}\right)+
\frac{A^{(2)}}{4\beta_{1}^{2}}\left(\frac{\hat{q}}{\hat{b}}-1\right)
-\left[\frac{A^{(2)}}{4\beta_{1}^{2}}-\frac{A^{(1)}}{4\beta_{1}}
\ln\left(\frac{e^{2\gamma_E-1}}{2}\right)\right]
\ln\left(\frac{\hat{q}}{\hat{b}}\right)
\nonumber \\
&&+\frac{A^{(1)}\beta_{2}}{4\beta_{1}^{3}}\hat{q}\left[
\frac{\ln(2\hat{q})+1}{\hat{q}}-\frac{\ln(2\hat{b})+1}{\hat{b}}\right]
+\frac{A^{(1)}\beta_{2}}{8\beta_{1}^{3}}\left[
\ln^{2}(2\hat{q})-\ln^{2}(2\hat{b})\right]
\nonumber \\
&&+\frac{A^{(1)}\beta_{2}}{8\beta_{1}^{3}}
\ln\left(\frac{e^{2\gamma_E-1}}{2}\right)\left[
\frac{\ln(2\hat{q})+1}{\hat{q}}-\frac{\ln(2\hat{b})+1}{\hat{b}}\right]
-\frac{A^{(1)}\beta_{2}}{16\beta_{1}^{4}}\left[
\frac{2\ln(2\hat{q})+3}{\hat{q}}-\frac{2\ln(2\hat{b})+3}{\hat{b}}\right]
\nonumber \\
& &-\frac{A^{(1)}\beta_{2}}{16\beta_{1}^{4}}
\frac{\hat{q}-\hat{b}}{\hat{b}^2}\left[2\ln(2\hat{b})+1\right]
+\frac{A^{(2)}\beta_{2}^2}{432\beta_{1}^{6}}
\frac{\hat{q}-\hat{b}}{\hat{b}^3}
\left[9\ln^2(2\hat{b})+6\ln(2\hat{b})+2\right]
\nonumber \\
&& +\frac{A^{(2)}\beta_{2}^2}{1728\beta_{1}^{6}}\left[
\frac{18\ln^2(2\hat{q})+30\ln(2\hat{q})+19}{\hat{q}^2}
-\frac{18\ln^2(2\hat{b})+30\ln(2\hat{b})+19}{\hat{b}^2}\right]
\label{sss}
\end{eqnarray}
\end{widetext}
where $\hat q$ and $\hat b$ are defined by
\begin{equation}
{\hat q} \equiv  {\rm ln}\left(xQ/(\sqrt 2\Lambda_{QCD})\right),~
{\hat b} \equiv  {\rm ln}(1/b\Lambda_{QCD})
\end{equation}
The coefficients $\beta_{i}$ and $A^{(i)}$ are
\begin{eqnarray}
& &\beta_{1}=\frac{33-2n_{f}}{12}\;,\;\;\;\beta_{2}=\frac{153-19n_{f}}{24}\; ,
A^{(1)}=\frac{4}{3}\;,
\nonumber \\
& & A^{(2)}=\frac{67}{9}-\frac{\pi^{2}}{3}-\frac{10}{27}n_
{f}+\frac{8}{3}\beta_{1}\ln\left(\frac{e^{\gamma_E}}{2}\right)\;
\end{eqnarray}
and $\gamma_E$ is Euler constant.

\section{\label{b}Light Meson Distribution Amplitudes}
The transverse-momentum-dependent light meson distribution amplitudes are
$\phi_{\pi}(x,k_{q\perp})$, $\phi_{\pi}^M(x,k_{q\perp})$, and $\phi^\sigma_{\pi}(x,k_{q\perp})$. The transverse-momentum-dependence
is assumed to be a Gaussian form and appears as a factorized part from the longitudinal wave functions.
Transformed into $b$-space, the distribution amplitudes can be written as \cite{wy2002}
\begin{equation}
    \phi(x,b)=\phi(x)\exp\left(-\frac{b^2}{4\beta^2}\right).
\end{equation}
Here, we denote the $b$-space distribution amplitudes as $\phi_{\pi}(x,b)$, $\phi_{\pi}^M(x,b)$, and $\phi^\sigma_{\pi}(x,b)$. As discussed previously in Ref. \cite{wang-yang2023}, see also Refs. \cite{wy2002} and \cite{JK93},
we adopt $\beta=4.0 \;\mathrm{GeV}^{-1} $ for the wave functions of pion, kaon, and $\eta_{q,s}$ mesons.
The expressions for the twist-2 and twist-3 distribution amplitudes are given by \cite{Ball-Braun2006}

\begin{equation}
\label{eq:phi}
\phi_\pi (x)=6x(1-x)\biggl[1+a_1^\pi C_1^{3/2}(t)+a_2^\pi C_2^{3/2}(t)\biggr],
\end{equation}
\begin{equation}
\label{eq:phip}
\begin{split}
\phi_\pi^P(x)&=1+a_{0P}^\pi+a_{1P}^\pi C_1^{1/2}(t)+a_{2P}^\pi C_2^{1/2}(t) \\
&\quad+a_{3P}^\pi C_3^{1/2}(t) +a_{4P}^\pi C_4^{1/2}(t) \\
&\quad+b_{1P}^\pi \ln(x)+b_{2P}^\pi \ln(1-x),\\
\end{split}
\end{equation}
\begin{equation}
\label{eq:phis}
\begin{split}
\phi'^\sigma_\pi(x)&=6x(1-x)\biggl[1+a_{0\sigma}^\pi+a_{1\sigma}^\pi C_1^{3/2}(t) \\
&\quad+a_{2\sigma}^\pi C_2^{3/2}(t)+a_{3\sigma}^\pi C_3^{3/2}(t)\biggr] \\
&\quad+9x(1-x)\biggl[b_{1\sigma}^\pi \ln(x)+b_{2\sigma}^\pi \ln(1-x)\biggr],\\
\end{split}
\end{equation}
where $t$ is defined as $t=2x-1$.
These $C$ functions are Gegenbauer polynomials.
The coefficients appearing in Eqs.~\eqref{eq:phi}--\eqref{eq:phis},
with $a_{i(P,\sigma)}^\pi $ for $i = 1,2,3,4$ and $b_{j(P,\sigma)}^\pi $ for $j =1,2$,
have the following values
\begin{equation}
  \begin{split}
    &a_1^\pi=0, \quad a_2^\pi=0.25\pm 0.15, \\
    &a_{0P}^\pi=0.048\pm 0.017, \quad a_{2P}^\pi=0.62\pm 0.21,\\
    &a_{4P}^\pi=0.089\pm 0.071, \quad a_{1P}^\pi=a_{3P}^\pi=0, \\
    &b_{1P}^\pi=b_{2P}^\pi=0.024\pm 0.009, \\
    &a_{0\sigma}^\pi=0.034\pm 0.014,\quad a_{2\sigma}^\pi=0.12\pm 0.05, \\
    &a_{1\sigma}^\pi=a_{3\sigma}^\pi=0, \quad b_{1\sigma}^\pi=b_{2\sigma}^\pi=0.016\pm 0.006. \\
\end{split}
\end{equation}

The parameters listed above are all determined at the renormalization scale of $\mu=1.0~\mathrm{GeV}$.
It is worth noting that, considering the similarity in quark composition between $\eta_{q,s}$ meson and pion,
we employ the same expressions for $\eta_{q,s}$ meson parameters as for the pion,
with appropriate substitutions made only for parts involving meson masses, quark masses, and decay constants.

For $ \rho $ meson, the twist-2 and twist-3 distribution amplitudes are given by \cite{PB-GWJ2007}
\begin{equation}
\phi_{\rho}(x)=6x(1-x)[1+a_{1}^{||} C_{1}^{3/2}(t)+a_{2}^{||}C_{2}^{3/2}(t)],
\end{equation}

\begin{eqnarray}
\phi_{\rho}^{t}(x)=&&3t^{2}+ \dfrac{3}{2}a_{1}^{\perp}t(3t^{2}-1)+\dfrac{3}{2}a_{2}^{\perp}t^{2}(5t^{2}-3)\nonumber \\
&&+(\dfrac{15}{2}\kappa_{3}^{\perp}-\dfrac{3}{4}\lambda_{3}^{\perp})t(5t^{2}-3)+\dfrac{5}{8}\omega_{3}^{\perp}(35t^{4}\nonumber \\
&&-30t^{2}+3)+\dfrac{3}{2}\dfrac{m_{q}+m_{q}}{m_{\rho}}\frac{f_{\rho}^{||}}{f_{\rho}^{\perp}}\\
&&\times \biggl\{1+ 8a_{1}^{||}t+ 3a_{2}^{||}[7-30x(1-x)]+t(1\nonumber \\
&&+3a_{1}^{||}+6a_{2}^{||}) \ln(1-x)
-t(1-3a_{1}^{||}+6a_{2}^{||})\ln x\biggr\}, \nonumber
\end{eqnarray}

\begin{eqnarray}
\Psi_{3}^{||}(x)=&&6x(1-x)\biggl[ 1+(\dfrac{1}{3}a_{1}^{\perp}+\dfrac{5}{3}\kappa_{3}^{\perp})C_{1}^{3/2}(t)\nonumber \\
&&+(\dfrac{1}{6}a_{2}^{\perp}+\dfrac{5}{18}\omega_{3}^{\perp})C_{2}^{3/2}(t) -\dfrac{1}{20}\lambda_{3}^{\perp}C_{3}^{3/2}(t)\biggr]\nonumber \\
&&+3\dfrac{m_{q}+m_{q}}{m_{\rho}}\dfrac{f_{\rho}^{||}}{f_{\rho}^{\perp}}\biggl\{x(1-x)[1+2a_{1}^{||}t\\
&&+3a_{2}^{||}(7-5x(1-x))]+(1+3a_{1}^{||}\nonumber \\
&&+6a_{2}^{||})(1-x)\ln(1-x)\nonumber \\
&&+(1-3a_{1}^{||}+6a_{2}^{||})x\ln x\biggr\},\nonumber
\end{eqnarray}
where $t=2x-1$, $C$ functions are Gegenbauer polynomials and $ \phi_{\rho}^{s}(x)=\dfrac{1}{2}\dfrac{d\Psi_{3}^{||}(x)}{dx} $. The coefficients in distribution amplitudes are as follows,

\begin{eqnarray}
&&a_{1}^{||}=0, \quad a_{2}^{||}=0.15\pm0.07, \nonumber \\
&&a_{1}^{\perp}=0, \quad a_{2}^{\perp}=0.14\pm0.06, \nonumber \\
&&\kappa_{3}^{\perp}=0, \quad \omega_{3}^{\perp}=0.55\pm0.25,\\
&&\lambda_{3}^{\perp}=0, \quad m_{q}=0.0056\pm0.0016\nonumber
\end{eqnarray}
The Gegenbauer polynomials are given by
\begin{eqnarray}
    &&C_1^{1/2}(t)=t,\nonumber \\
    &&C_2^{1/2}(t)=\frac{1}{2}\left(3t^2-1\right), \nonumber \\
    &&C_3^{1/2}(t)=\frac{t}{2}\left(5t^2-3\right), \\
    &&C_4^{1/2}(t)=\frac{1}{8}\left(35t^4-30t^2+3\right), \nonumber
\end{eqnarray}
and
\begin{eqnarray}
    &&C_1^{3/2}(t)=3t, \nonumber \\
    &&C_2^{3/2}(t)=\frac{3}{2}\left(5t^2-1\right), \nonumber \\
    &&C_3^{3/2}(t)=\frac{5}{2}t\left(7t^2-3\right),  \\
    &&C_4^{3/2}(t)=\frac{15}{8}\left(21t^4-14t^2+1\right). \nonumber
\end{eqnarray}

\end{document}